\documentclass[12pt,a4paper]{article}
\usepackage{moreverb}
\usepackage{url}
\usepackage{bm}
\usepackage{grffile}
\usepackage{newtxmath}
\usepackage{natbib}
\usepackage{graphicx}
\usepackage{lineno}

\usepackage{amsmath}
\usepackage{algorithm}
\usepackage[UKenglish]{isodate}
\usepackage{algorithm}
\newcommand{\cC}{\mathcal{C}}

\usepackage[usenames,dvipsnames]{xcolor}

\newcommand{\commentF}[1]{#1}  
\newcommand{\commentR}[1]{#1}   
\newcommand{\commentBG}[1]{#1}   

\newcommand{\by}{\mathbf{y}}

\newcommand\BibTeX{{\rmfamily B\kern-.05em \textsc{i\kern-.025em b}\kern-.08em
		T\kern-.1667em\lower.7ex\hbox{E}\kern-.125emX}}
\newcommand\MiKTeX{{\rmfamily M\kern-.05em \textsc{i\kern-.025em K}\kern-.08em
		T\kern-.1667em\lower.7ex\hbox{E}\kern-.125emX}}
\newcommand\PracTeX{{\rmfamily P\kern-.05em \textsc{r\kern-.025em a\kern-.025em
			c}\kern-.08em
		T\kern-.1667em\lower.7ex\hbox{E}\kern-.125emX}}

\begin{document}
\cleanlookdateon
\title{Spying on the prior of the number of data clusters and the partition distribution in Bayesian cluster analysis}
\author{Jan Greve\thanks{WU Vienna University of
		Business and Economics},
	Bettina Gr\"un\thanks{WU Vienna University of Business and
		Economics},	
	Gertraud Malsiner-Walli\thanks{WU
		Vienna University of Business and Economics}~{}and\\
	Sylvia Fr\"uhwirth-Schnatter\thanks{WU Vienna University of
		Business and Economics}
	\\\vspace{0.5cm}}

\date{} 

\maketitle

\sloppy
\begin{small}
\begin{abstract}
\commentBG{Cluster analysis aims at partitioning data into groups or clusters. In applications, it is common to deal with problems where the number of clusters is unknown}. Bayesian mixture models employed in such applications usually specify a flexible prior that takes into account the uncertainty with respect to the number of clusters. However, a major empirical challenge involving the use of these models is in the characterisation of the induced prior on the partitions. This work introduces an approach to compute descriptive statistics \commentBG{of the prior on the partitions} for three selected Bayesian mixture models developed in the areas of Bayesian finite mixtures and Bayesian nonparametrics. The proposed methodology involves computationally efficient enumeration of the prior on the number of clusters in-sample (termed as ``data clusters'') and determining the first two prior moments of symmetric additive statistics characterising the partitions. The accompanying reference implementation is made available in the \textsf{R} package \textbf{fipp}. Finally, we illustrate the proposed methodology through comparisons and also discuss the implications for prior elicitation in applications.
\end{abstract}
\end{small}
%
%

\section{Introduction}
\label{sec:intro}

Methodologically, cluster analysis aims at partitioning observations into a set of mutually exclusive groups such that observations within the same group share some characteristics and are differentiable from observations across other groups. In model-based clustering, this problem is commonly dealt with using mixture models where data are assumed to be drawn from a distribution whose density is specified as a convex combination of parametric densities referred to as mixture components. In the context of clustering, the most natural understanding of mixture components is that each of them represents a distinct group within the population, \commentF{see \cite{McLachlan+Peel:2000} and \cite{gru:mod} for a recent review}.

\commentF{However,
 the number} of distinct groups within the population is often not known a-priori. Therefore, mixture models are often used solely as flexible modelling tools where the components are not necessarily associated with any known or observed quantity of inferential importance.
Nonetheless, in the Bayesian mixture framework \commentF{(see \citealt{fru:book} for a comprehensive review),} the data generating process where mixture components represent potential groups present in the population can be specified regardless of whether any information about this quantity is available a-priori or not.  By assigning a prior distribution to this quantity, the uncertainty with respect to this variable can be reflected in the model. This fully Bayesian approach to the clustering problem with an unknown number of mixture components was proposed by \cite{Richardson+Green:1997}. Furthermore, the link between this approach and
 \commentF{Bayesian Nonparametric (BNP) mixtures (see, e.g., \citealt{lij-pru:mod})} has been recently explored by \cite{Miller+Harrison:2018}, who coined the term Mixture of Finite Mixtures (MFM\commentBG{s}) to refer to the aforementioned approach, \commentF{and by \cite{Fruehwirth-Schnatter+Malsiner-Walli+Gruen:2020}}.

In addition, \commentBG{Bayesian methods enable the clear distinction of} clusters realised in the sample and those in the population. Specifically, mixture components associated with at least one observation belong to the former category and the latter \commentBG{category} consists of components with no observations. This distinction is usually not made in the frequentist framework that utilises \commentF{maximum likelihood (ML)} estimation. Details of the application and estimation of finite mixtures in \commentBG{a} ML framework are provided in \cite{McLachlan+Peel:2000}. \commentF{Both ML and Bayesian methods are reviewed in \cite{fru-etal:han}.}


Although theoretically sound and intuitively straightforward, one major empirical challenge of Bayesian mixture models with an unknown number of components \commentF{lies} in the prior and hyperparameter specification. This applies to MFM models developed in the area of Bayesian finite mixtures as well as \commentF{to} infinite mixture models originating in the BNP literature. All these models need to be specified with a hyperparameter (with or without a hyperprior) assigned to the prior distribution on the mixture component weights. In addition, for MFM models in particular, a prior on the number of mixture components needs to also be specified. Crucially, these specifications implicitly induce a prior distribution on the partitions of the data. The empirical importance of studying this prior can be easily understood as characteristics of the posterior \commentBG{of the} partitions with substantial inferential importance -- the number of clusters in the sample and allocation of data to clusters -- are directly influenced by this prior.

Hence, this work aims at complementing applied works involving Bayesian mixture models with an unknown number of components by providing methods to quantify \commentF{the prior distribution}
on the partitions via ``spying'' on its probabilistic characteristics. To facilitate this goal, it introduces a way to evaluate moments of certain types of statistics defined on the induced prior on \commentBG{the} partitions. Specifically, for three important Bayesian finite and infinite mixture models, we present formulas for the  prior on the number of \commentBG{data} clusters induced by the prior on the partitions and the first two moments of any symmetric additive functionals defined over the prior partitions. In addition, we derive computationally feasible evaluations of these quantities and provide a reference implementation written in \textsf{R} \citep{R-Core-Team:2020} in the package \textbf{fipp} \citep{Greve:2021}. The three models in question are the Dirichlet Process Mixture (DPM) model by \cite{Ferguson:1973}, the \commentBG{MFM} model proposed by \cite{Miller+Harrison:2018} and its generalisation by \cite{Fruehwirth-Schnatter+Malsiner-Walli+Gruen:2020}. These are referred to as the DPM, the static MFM and the dynamic MFM following the naming convention used in \cite{Fruehwirth-Schnatter+Malsiner-Walli+Gruen:2020}. To demonstrate the practical relevance of the methodology, a juxtaposition of these two MFM models is made based on their characteristics of the induced prior on the partitions. Additionally, an empirical comparison between these three models under popular prior and hyperparameter settings is also conducted.

This paper is structured as follows: Section~\ref{sec:mixt-models-bayes} reviews the different mixture models considered in this paper for Bayesian cluster analysis. The explicit priors used in Bayesian mixture models which give rise to the aforementioned three models are discussed in Section~\ref{sec:explicit-priors}. The main contribution of this work \commentBG{is to facilitate} quantification \commentBG{and characterisation} of the induced priors on the partitions. Section~\ref{sec:induced-priors} \commentBG{derives the theoretic results} together with computationally feasible algorithms for the calculation. Additionally, it positions the contribution of this work relative to previous works in this area both in theory and in applications. In Section~\ref{sec:insp-induc-priors}, we investigate the differences between the static and dynamic MFMs using the tools developed in the previous section. In Section~\ref{sec:comp-defa-priors} we empirically compare the implicit prior distributions for default prior specifications suggested in Bayesian cluster analysis.  Section~\ref{sec:conclusions} outlines how these prior considerations might be used for prior elicitation in an application, and finally a summary of our findings is provided at the end in Section~\ref{sec:summary}.

\section{Mixture models for Bayesian cluster
analysis}\label{sec:mixt-models-bayes}
One useful way to represent certain types of Bayesian mixture models employed in cluster analysis is to formulate them as generative models involving random partitions $\cC$. This is a common approach in the literature of product partition models, see for example \cite{Hartigan:1990} and \cite{Barry+Hartigan:1992}. In general, Bayesian mixture models based on exchangeable random partitions (for detailed coverage of this topic, see \commentF{\citealt{pit:com}}) can be reformulated in this way. All three models (the static and dynamic MFM models and the DPM) considered in this work fall under this category and thus can be written \commentF{hierarchically, involving  random partitions $\cC$}.

Consider a partition $\cC$ that separates $N$ observations with observed responses $\{\by_1, \ldots, \by_N\}$ by grouping the set of indices of the data $[N] \coloneqq \{1,\ldots,N\}$. Such a partition $\cC= \{\cC_1,\ldots,\cC_{K_+}\}$ consists of blocks $\cC_k$, $ k = 1,\ldots,K_+$, where each block $\cC_k$ is a non-empty and disjoint subset of $[N]$ whose union is $[N]$. Hence, $\cC_k$ is interpreted as a cluster containing the indices of the observations assigned to it. Consequently, $K_+ = \lvert \cC \rvert$, the number of blocks in the partition, is interpreted as the number of clusters in the sample of $N$ observations. We refer to these realised clusters as ``data clusters\commentBG{''} as in \cite{Fruehwirth-Schnatter+Malsiner-Walli+Gruen:2020} from here on. The resulting Bayesian mixture model based on $\cC$ has a representation
\begin{linenomath*}
\begin{align}
\commentF{\cC}  &\sim p(\cC), &\nonumber\\
\commentF{\bm{\theta}_k} &\sim p(\bm{\theta}_k),  &\text{\commentF{independently for }}
\commentF{k = 1, 2, \ldots },\label{eq:genmodelpartitions}\\
\commentF{\by_i|i \in \cC_k, \bm{\theta}_k}& \sim
f(\by_i|\bm{\theta}_k),  & \text{\commentF{independently for }} \commentF{i=1, \ldots, N,} \nonumber
\end{align}
\end{linenomath*}
\commentF{where $f$ is the density function of the distribution the observations are assumed to be drawn from and $\bm{\theta}_k$ is the set of parameters specific to the density function $f$ of the subgroup $k$.}

In Bayesian cluster analysis, such a partition $\cC$  is \commentBG{usually} induced by a sequence of random variables, i.e., a vector of categorical variables called class assignment vector $\bm{S} = (S_1,\ldots,S_N)$. Each element $S_i\in \{1,\ldots,K\}$, $ i = 1,\ldots,N,$ takes one of the $K$ class labels as its realisation with $K$ being the number of components in the mixture distribution. Therefore, clustering arises in a natural way as each block $\cC_k,\ k = 1,\ldots, K_+$, within the partition $\cC$ is induced by $\bm{S}$ by grouping $S_i = S_j,\ i\neq j$ for all $i$ and $j$ in $[N]$, \commentF{see e.g. \cite{lau-gre:bay}}.
Hence, amongst class labels of $S_i$ ranging from 1 to $K$, only $K_+$ unique labels are present in $\bm{S}$ which induce $\cC = \{\cC_1,\ldots,\cC_{K_+}\}$.

Typically in Bayesian finite mixture models, a $K$-variate symmetric Dirichlet-Multinomial distribution is considered as a prior on the class assignment vector $\bm{S}$. MFM models considered in this work all use the Dirichlet-Multinomial distribution as a partition generator (for other choices  of the prior, see  for example \citealt{Lijoi+Pruenster+Rigon:2020}). Additionally, in the MFM framework, a prior on $K$ is specified, usually a discrete distribution on $\mathbb{N}_+$. Hence, the MFM models considered in this work have the following generative model for the partitions:
\begin{linenomath*}
\begin{align}
K &\sim p(K),\nonumber\\
\bm{\eta}_K | K, \gamma_K &\sim
\mathcal{D}_K(\gamma_K),\label{eq:dirichletmultinomial}\\
S_i | \bm{\eta}_K &\sim \mathcal{M}_K(1, \bm{\eta}_K),\qquad \text{for } i=1,\ldots,N,\nonumber
\end{align}
\end{linenomath*}
where $p(K)$ is the aforementioned discrete prior on $K$. Here, the Dirichlet-Multinomial prior is written hierarchically with the $K$-variate symmetric Dirichlet prior $\mathcal{D}_K$ on the weight vector $\bm{\eta}_K = (\eta_1,\ldots,\eta_K)$ followed by the $K$-variate Multinomial prior $\mathcal{M}_K$ on the class assignment vector $\bm{S}$. Crucially, the characteristics of the induced \commentBG{prior on the} partitions of this model \commentBG{are} determined by the choice of $p(K)$ and the Dirichlet parameter $\gamma_K$ as well as the sample size $N$.

Finally, conditional on the resulting class assignments $\bm{S}$, the likelihood of each $\by_i$ is evaluated as follows:
\begin{linenomath*}
\begin{align}
    p(\by_i|S_i = k,\commentF{\bm{\theta}_1,\ldots,\bm{\theta}_{K}}) = f(\by_i|\commentF{\bm{\theta}_k}).\label{emission-cond-Si}
\end{align}
\end{linenomath*}
%
 \commentF{The} widely known mixture density  conditional on the number of mixture components $K$, the component weight vector $\bm{\eta}_K$ and the component specific parameter vector $\Theta_K =  (\bm{\theta}_1,\ldots,\bm{\theta}_K)$ is obtained through integrating out the aforementioned class assignments $\bm{S}$:
\begin{linenomath*}
\begin{align}
  p(\by_i|K,\Theta_K,\bm{\eta}_K)&= \sum_{k=1}^K \eta_k f(\by_i |\commentF{\bm{\theta}_k} ).\label{emission-uncond-Si}
\end{align}
\end{linenomath*}
Note that there is a crucial distinction between $K$, the number of components in the mixture distribution, and $K_+$, the number of clusters in the data (i.e., \commentBG{the} data clusters). While $K$ is assumed to represent the number of clusters in the population, $K_+$ can be interpreted as the number of \commentBG{clusters out of the $K$ clusters in the population} that generated the data at hand. Therefore, of all $K$ mixture components in Equation~\eqref{emission-uncond-Si}, only $K_+$ densities are associated to at least one observation $\by_i$ through $S_i$ as in Equation~\eqref{emission-cond-Si}. \commentBG{Given} this definition of $K_+$, as the finite-sample characteristics of $K$, its upper bound is defined as $K_+\leq \min(K,N)$.

It is evident that the prior on the partitions is not influenced by the choice of the density function $f$, nor its parameters $\Theta_K$. Since the aim of this work is to quantify the characteristics of the induced prior on the partitions, the focus will exclusively be on the explicit and implicit characteristics of Equation~\eqref{eq:dirichletmultinomial} from here on.
\section{Explicit prior}\label{sec:explicit-priors}
The main objective of this study is to characterise the \commentBG{prior on the} partitions induced by the model given in Equation~\eqref{eq:dirichletmultinomial}. Once $p(K)$ and $\gamma_K$ are specified, the prior distribution on the partitions of $[N]$ is completely determined. In this section, we consider several modelling approaches for clustering previously explored in the literature of Bayesian Finite Mixtures and Bayesian Nonparametrics by focusing solely on their specification of $p(K)$ and $\gamma_K$.
\subsection{Prior hyperparameter on the weight distribution}\label{sec:prior-hyper-weight}
Conditional on \commentBG{a} given $K$, a $K$-variate symmetric Dirichlet distribution $\mathcal{D}_K(\gamma_K)$ on the weights $\bm{\eta}_K=(\eta_1,\ldots,\eta_K)$ is specified by choosing a hyperparameter $\gamma_K$. The static and dynamic MFM models coined by \citet{Fruehwirth-Schnatter+Malsiner-Walli+Gruen:2020} refer to two MFM models which differ only in the form of this hyperparameter in the following way:
\begin{linenomath*}
\begin{align*}
&\text{static MFM: }\qquad \gamma_K\equiv \gamma,\\
&\text{dynamic MFM: }\quad \gamma_K = \frac{\alpha}{K}.
\end{align*}
\end{linenomath*}
That is, the Dirichlet parameter $\gamma_K$ is fixed to a constant $\gamma$ regardless of $K$ for the static MFM, while that of the dynamic MFM is inversely proportional to $K$ with the specific form of $\alpha/K$.

The induced prior on the partitions differs considerably across these two specifications, as already noted by \cite{McCullagh+Yang:2008}. To put it simply, for larger values of $K$, the Dirichlet parameter $\gamma_K$ of the dynamic MFM approaches zero, thus preferring a sparse distribution of $\bm{\eta}_K$.
This implies that the larger $K$, the more likely it is that $K_+$ is smaller than $K$ causing an increasing gap between the number of components and data clusters. In fact, the DPM can be considered a limiting case of this dynamic MFM where $K$ is taken to infinity \citep{Green+Richardson:2001}, thus with probability one $K_+$ is less than $K$. For this reason, the parameter $\alpha$ in the DPM corresponds to the parameter $\alpha$ in the dynamic MFM. On the other hand, for the static MFM, the difference between $K_+$ and $K$ depends more heavily on the value of $\gamma$.

\subsection{Prior on $K$}\label{sec:prior-K}
For both MFM models, the prior on $K$ is usually given a proper discrete distribution with support on $\mathbb{N}_+$ to ensure the posterior on $K$ to also be proper \citep{Nobile:2004}. The DPM being a limiting case of the dynamic MFM uses a degenerate prior on $K$ with a point mass on infinity. Some other choices of $p(K)$ previously proposed in the literature we consider in later sections are: the uniform prior on $K$ between 1 and 30 proposed by \citet{Richardson+Green:1997}, the geometric prior \commentF{$\mbox{\rm Geo}(0.1)$} 
on $K-1$ suggested in \cite{Miller+Harrison:2018} and the beta-negative-binomial prior $\text{BNB}(1,4,3)$ on $K-1$ considered in \citet{Fruehwirth-Schnatter+Malsiner-Walli+Gruen:2020}. The uniform prior is an example where the support on $p(K)$ is not on $\mathbb{N}_+$, but rather \commentBG{is on a} truncated domain. The other two priors share the characteristics of having a monotonically decreasing probability mass function, thus penalising additional components a-priori. 
\section{Induced prior on the partitions}\label{sec:induced-priors}

The model specification for generating partitions outlined in Equation \eqref{eq:dirichletmultinomial} explicitly characterises the number of mixture components $K$ and the class assignments $S_i,\ i = 1,\ldots N$, as well as the intermediate weight vector $\bm{\eta}_K$. A partition $\cC$ is then induced from the sampled class assignment vector $\bm{S} = (S_1,\ldots,S_N)$. While this way of hierarchically combining well known probability distributions to generate partitions is implementationally straightforward, it masks the actual
\commentF{prior distribution on $\cC$}. In other words, \commentF{Equation~\eqref{eq:dirichletmultinomial}} is not particularly informative in understanding its finite-sample characteristics conditional on the given $N$ which is the prior distribution on $\cC$. In clustering, however, it is particularly important to understand the finite-sample characteristics of the model as the prior information that directly influences the clustering behaviour are not the latent quantities such as $K$ and $\bm{\eta}_K$, but rather the realised characteristics of the mixture distribution such as the partitions. Studying the prior on the partitions can therefore be crucial for many reasons. E.g., one may like to evaluate the informativeness of the induced prior on the partitions relative to the resulting posterior partitions to ensure that proper learning from the data took place. Also, in some applications, one may want to incorporate external information into the prior partitions concerning the ``kind'' of partitions one is interested in, e.g.~regarding the assumed number of clusters $K_+$ in the data. This information however, cannot be directly embedded in the model as neither $p(K)$ nor $\gamma_K$ will single-handedly control the prior on the partitions.

Hence, it is of paramount importance to quantify the induced prior on the partitions so as to ``spy'' on its characteristics. For this reason, this section deals with delineating all the steps and procedures that enable characterisation of the induced prior on the partitions. Specifically, for all three models outlined in Section \ref{sec:prior-hyper-weight}, two \commentBG{possibilities to characterise this prior} are \commentBG{considered}: the prior distribution on $K_+$, and the first two \commentBG{prior} moments of any symmetric additive functional defined over the partitions conditional on $K_+$. Finally, \commentBG{combining these quantities enables determining the prior moments of these functionals} unconditional on $K_+$.

Characterisation of \commentBG{the} probability distribution on the partitions is an arduous task due to its combinatorial construction. \commentF{\cite{Gnedin:2010}} mentions the use of expectations of symmetric statistics computed over the exchangeable frequency vector (i.e., \commentBG{the} normalised block sizes). This relates to the first moment of the symmetric additive functionals unconditional on $K_+$ that this paper introduces as one of the descriptive statistics for the characterisation of the \commentBG{prior on the} partitions. In addition, this paper also considers a way to \commentBG{quantify the variability of these functionals} via the corresponding variance and also derives a way to compute these quantities efficiently and provides a reference implementation in the \textsf{R} package \textbf{fipp}.

Note that the \commentBG{\textsf{R}} package \textbf{AntMAN} \citep{Argiento:2019} \commentBG{also} allows the evaluation of the prior  on $K_+$ for the static MFM. However, \textbf{fipp} offers more comprehensive tools for the characterisation \commentBG{of the prior on the partitions for} the DPM and \commentBG{the static and dynamic} MFM models. \commentBG{In particular, package \textbf{fipp} also provides the} capacity to evaluate \commentBG{the prior} expectation and variance of any symmetric additive functional defined over the  partitions.


\subsection{The induced EPPF}\label{sec:induced-eppf}

The induced prior on the partitions is available for all three modelling approaches: the DPM, the static MFM and the dynamic MFM. All these priors are symmetric functions of the data cluster sizes $\{\lvert \cC_1\rvert,\ldots, \lvert \cC_k\rvert\}$ for $K_+  =k$ and hence, $p(\cC |N, \bm{\gamma})$ with $\bm{\gamma} = \{\gamma_K\}$ is an exchangeable partition probability function (EPPF) in the sense of \citet{Pitman:1995} and defines an exchangeable random partition of the $N$ data points for all three classes of mixture models.

For a DPM with  concentration parameter $\alpha$, the EPPF on a partition $\cC = \{\mathcal{C}_1, \ldots, \mathcal{C}_k\}$ with $K_+ = k$ is given as the following Ewens distribution:
\begin{linenomath*}
\begin{align*}
    p(\mathcal{C} | N, \alpha )=
 \frac{ \alpha
^{k} \Gamma(\alpha) }{
  \Gamma(\alpha + N)} \prod_{j=1}^{k} \Gamma(\lvert \cC_j \rvert),
\end{align*}
\end{linenomath*}
Similarly, for a static MFM with $\gamma_K \equiv \gamma$ and thus conditional only to $\gamma$ rather than the sequence $\bm{\gamma}$, the EPPF of the same $\cC$ is given in \citet{Miller+Harrison:2018} as follows:
\begin{linenomath*}
\begin{align*}
p(\mathcal{C}   | N,  \gamma )  &=    {V} ^\gamma _{N, {k}}  \prod_{j=1}^{k}
\frac{ \Gamma(\lvert \cC_j\rvert + \gamma)}{ \Gamma(\gamma) } , \\
{V}^\gamma _{N, {k}} &=  \sum_{K=k}^\infty
p(K)  \frac{K !}{(K- k )!} \frac{\Gamma(\gamma K) }{  \Gamma(\gamma K + N)} ,
\end{align*}
\end{linenomath*}
\commentF{as proven earlier by \cite{Gnedin+Pitman:2006} in the BNP
  literature.  ${V}^\gamma_{N,k}$ (related to the $V$-weights
  $\tilde{V}^\gamma_{N,k}$ in \citealt{Gnedin+Pitman:2006} through the
  normalisation $ \tilde{V}^\gamma_{N,k} = \gamma ^k {V} ^ \gamma _{N,k} $)
  is} associated to the across block characteristics of the partition
$\cC$ such as $\commentBG{k}$ and $N$, as well as the hyperparameter $\gamma$
and the prior $p(K)$. This quantity can be computed recursively using
\citet[Proposition~3.2]{Miller+Harrison:2018}.\footnote{Note the
  following change of notation: $ V_{N,k} ^\gamma \equiv V_n(t)$ in
  \citet{Miller+Harrison:2018}.}  For $k=1, 2, \ldots$:
\begin{linenomath*}
\begin{align*}
  {V}^\gamma_{N+1,k+1}  &=  \frac{1}{ \gamma} {V} ^\gamma _{N,k} -  \left(
\frac{N}{\gamma} + k\right)  {V} ^\gamma _{N+1,k}  ,  \quad  {V} ^\gamma
_{{N},0}
          =   \sum_{K=1}^\infty  \frac{\Gamma(\gamma K) }{ \Gamma(\gamma  K +
N)} p(K).
\end{align*}
\end{linenomath*}
A generalisation of the above result is given in \citet{Fruehwirth-Schnatter+Malsiner-Walli+Gruen:2020}. Specifically, they consider an arbitrary sequence $\bm{\gamma} = \{\gamma_K\}$ for $K = \commentBG{1},\ldots,\infty$ assuming the prior on $K$ has support on $\mathbb{N}_+$. \commentBG{They refer to the MFM model with such an arbitrary sequence $\bm{\gamma}$ as the \emph{generalised MFM} model}. Thus the generalised EPPF of the same $\cC$ is written as follows:
\begin{linenomath*}
\begin{align}
\label{mixparti}
p(\mathcal{C} |N, \bm{\gamma})
 &=  \sum_{K=k}^\infty  p(K)   p(\mathcal{C} | N, K, \gamma_K  ),\\
  p(\mathcal{C} |  N, K,  \gamma_K )  &= V_{N, k}^{K, \gamma_K}
\prod_{j=1}^{k}  \frac{\Gamma(\lvert \cC_j\rvert +\gamma_K)}{\Gamma(\gamma_K)}
, \label{mixpartiK}
\\
\label{mixpKV}
V_{N,k }^{K, \gamma_K}   &=  \frac{ \Gamma(\gamma_K K) K ! }{ \Gamma(\gamma_K
K+N) (K- k )!} .
\end{align}
\end{linenomath*}
For the general\commentBG{ised} MFM, the $V$-weight also depends on $K$ as evident from Equation \eqref{mixpKV}. The explicit form of the EPPF for the dynamic MFM is obtained by setting $\gamma_K = \alpha / K$.

\subsection{The induced prior on the number of data clusters
$K_+$}\label{sec:induced-prior-number}

The prior $p(K_+| N, \bm{\gamma})$ on the number of data clusters $K_+$, where the uncertainty with respect to $K$ is integrated out and one accounts for the specification of $\bm{\gamma} = \{\gamma_K\}$ and the sample size $N$, could be derived from the EPPF given in Equation~\eqref{mixparti} by summing over all partitions $\mathcal{C}$ with $K_+ = k$ data clusters across all $k = 1,\ldots,N$. A naive approach would be to sum over the set of all partitions of $[N]$ with $K_+ = k$ clusters for all $k$ which amounts to a computation in the order of the $N$-th Bell number $B_N$ (for details, see Appendix~\ref{sec:EPPF-support-combinatoric-part}).

An alternative approach to obtain $p(K_+| N,\bm{\gamma})$ is suggested in \citet[\commentF{Theorem~3.1}]{Fruehwirth-Schnatter+Malsiner-Walli+Gruen:2020}.
They base the derivation on the prior $p(N_1, \ldots, N_{K_+} |N,\bm{\gamma})$ using $(N_1,\ldots,N_{K_+})$ which they call {\em labelled} data cluster sizes where cluster sizes are arranged in some exchangeable random order resulting in labels $\{1,\ldots, K_+\}$ being attached to the $K_+$ data clusters in $\mathcal{C}$ as $\lvert \cC_j \rvert = N_j, j = 1,\ldots, K_+$. By this operation, all class assignment vectors $\bm{S}$ equivalent under the exchangeability are mapped into a set partition $\cC = \{\cC_1,\ldots,\cC_{K_+}\}$ by $\binom{K}{K_+}K_+!$-to-one mapping while $\bm{S}$ to $(N_1,\ldots,N_{K_+})$ is $\binom{K}{K_+}\binom{N}{N_1N_2\cdots N_{K_+}}$-to-one. Hence, the multiplicity of $(N_1,\ldots,N_{K_+})$ relative to $\cC$ is $\frac{1}{K_+!}\binom{N}{N_1N_2\cdots N_{K_+}}$ resulting in the EPPF on the labelled data cluster sizes as given by:
%
\begin{linenomath*}
\begin{align*}
    p(N_1,\ldots,N_{K_+}|N,K,\bm{\gamma}) = \frac{N!}{K_+!}\frac{ {V}_{N, K_+}^{K,
\gamma_K}}{ \Gamma(\gamma_K) ^{K_+}}\prod_{j=1}^{K_+}\frac{ \Gamma(N_{j} +\gamma_K)} {\Gamma(N_j + 1)}.
\end{align*}
\end{linenomath*}
Marginalising out $K$ leads to:
\begin{align*}
    p(N_1,\ldots,N_{K_+}|N,\bm{\gamma}) = \frac{N!}{K_+!}\sum_{K=K_+}^{\infty}p(K)\frac{ {V}_{N, K_+}^{K,
\gamma_K}}{ \Gamma(\gamma_K) ^{K_+}}\prod_{j=1}^{K_+}\frac{ \Gamma(N_{j} +\gamma_K)} {\Gamma(N_j + 1)},
\end{align*}
Then, summing up the probabilities of all labelled data cluster sizes $(N_1,\ldots,N_{k})$ with $K_+ = k$ amounts to computing $P(K_+ =k|N,\bm{\gamma})$. Thus, we have:
\begin{align}
  P(K_+ = k |N, \bm{\gamma}) &= \frac{N!}{ k!}
                               \sum_{K=k}^\infty p(K) \frac{ {V}_{N, k}^{K,
\gamma_K}}{ \Gamma(\gamma_K) ^{k}
                               } C^{K, \gamma_K}_{N,k},\label{pKplus}\\
 C^{K, \gamma_K}_{N,k} &=
  \sum_{\substack{N_1,\ldots, N_k > 0\\N_1+\ldots+N_k=N}}
  \prod_{j=1}^k \frac{ \Gamma(N_{j} +\gamma_K)} {\Gamma(N_j + 1)}\label{PposCk}.
\end{align}
where the term $C^{K, \gamma_K}_{N,k}$ sums over all possible labelled data cluster sizes $(N_1,\ldots,N_k)$.
As shown in \citet[\commentF{Algorithm~1}]{Fruehwirth-Schnatter+Malsiner-Walli+Gruen:2020}, $C_{N,k} ^{K, \gamma_K}$ can be determined recursively (see also Algorithm~\ref{KNMFM} in Appendix~\ref{sec:algor-comp-prior}).  For a static MFM, $C_{N,k}^{K, \gamma_K} \equiv C ^\gamma _{N,k}$ is independent of $K$ and can be obtained in a single recursion from Equation~\eqref{recck} in Appendix~\ref{sec:algor-comp-prior}. For a DPM, $w_n=1/n$ is used in recursion~\eqref{recck} in Appendix~\ref{sec:algor-comp-prior} to obtain $C^{\infty}_{N,k}$.

In principle, to determine the prior on the number of data clusters $K_+$, an infinite sum over $K$ has to be computed. Practically, a maximum value for $K$ is set to determine the prior. The missing mass is reflected by the prior on the number of data clusters $K_+$ not having a total mass of 1. Thus the total mass of the truncated prior covered can be used to check the suitability of the selected maximum value of $K$. If the mass of the truncated prior is assessed to be not sufficiently close to 1, the maximum value may be increased for a better approximation.

\subsection{The induced prior on the partitions based on the labelled data cluster sizes}\label{sec:char-prior-part}

The prior $p(K_+|N,\bm{\gamma})$ shown in Section~\ref{sec:induced-prior-number} can be considered a facet of the induced prior on the partitions $\cC$ that only concerns the number of blocks within each $\cC$ (which corresponds to the number of data clusters $K_+$ as explained in Section~\ref{sec:mixt-models-bayes}) while ignoring other characteristics. A generalisation of this approach is to consider a distribution of functionals defined over the induced prior on the partitions $\cC$. However, unlike the special case $p(K_+|N,\bm{\gamma})$, most functionals do not allow for easy derivation of such a distribution. Nevertheless, moments of some functionals with certain characteristics can be computed over the induced prior on the partitions conditional on the number of data clusters $K_+$. Specifically, we consider functionals defined over the labelled data cluster sizes $(N_1,\ldots,N_{K_+})$ which are symmetric and given as additive sums of functions of the single data cluster size $N_j$ \commentBG{over} all $j = 1,\ldots,K_+$. For these functionals we show that at least the first two moments can be easily derived and evaluated efficiently conditional on $K_+ \commentBG{=k}$:
\begin{linenomath*}
\begin{align*}
  \Psi(N_1,\ldots,N_{\commentBG{k}}) &= \sum_{j=1}^{\commentBG{k}} \psi(N_j).
\end{align*}
\end{linenomath*}
Relatedly, the aforementioned \commentF{\cite{Gnedin:2010}} considers an approach to compute expected values of the same statistics defined over \commentBG{the} exchangeable frequency vector without conditioning on $K_+$. However, combined with the distribution on $K_+$ derived in Section~\ref{sec:induced-prior-number}, the first two conditional moments we derive can also trivially be made unconditional on $K_+$.

For all three models considered in this work, the subsequent Section~\ref{sec:induc-cond-prior-1} introduces the prior on the labelled data cluster sizes $(N_1,\ldots,N_k)$ conditional on $K_+ = k$. This distribution is marginalised in Section~\ref{sec:marg-prior-labell} to obtain the conditional distribution on $N_j$ for all $j = 1,\ldots,k$ which is then used in Section~\ref{sec:comp-prior-moments} to evaluate the expectation of $\psi(N_j)$ and $\psi(N_j)\psi(N_l), j\neq l$ for all $j,l = 1,\ldots,k$. Based on these quantities, Section~\ref{sec:computing-prior-mean} derives the prior mean and variance of $\Psi(N_1,\ldots,N_{\commentBG{k}})$ conditional on $K_+\commentBG{=k}$. Furthermore, several functionals of empirical relevance are introduced as examples of symmetric additive statistics $\Psi(N_1,\ldots,N_{K_+})$. Finally, in Section~\ref{sec:computing-prior-mean-uncond} the prior mean and variance conditional on $K_+$ are combined with the prior distribution on $K_+$ to marginalise out $K_+$. In this way, we show that the first two moments of symmetric additive functionals defined over the labelled data cluster sizes can be computed unconditional on $K_+$ and in fact be evaluated rather efficiently in terms of computation by utilising recursion.

\subsubsection{The induced conditional prior on the labelled data cluster
sizes}\label{sec:induc-cond-prior-1}

The prior distribution $p(N_1, \ldots, N_{K_+}|N, \bm{\gamma})$ of the labelled data cluster sizes is defined over {\em all possible composition\commentBG{s} of $N$}, with $K_+$ being a random number taking a value $K_+= 1, \ldots,N$.  As pointed out by \citet{Green+Richardson:2001}, it is also interesting to consider the induced prior distribution over the labelled data clusters sizes for a given number of data clusters $K_+ = k$. This leads to the {\em conditional} prior on the labelled data cluster sizes for a given number of data clusters $K_+=k$ which is defined as:
\begin{linenomath*}
\begin{align*}
p(N_1, \ldots, N_k |N, K_+ = k, \bm{\gamma})&= \frac{p(N_1, \ldots, N_k|N,
\bm{\gamma}) }
{ P(K_+ = k |N , \bm{\gamma})},
\end{align*}
\end{linenomath*}
where $P(K_+ = k |N , \bm{\gamma})$ is the prior on the number of data clusters. \citet{Miller+Harrison:2018} provide this conditional prior for the DPM and derive it for the static MFM. In addition \citet{Fruehwirth-Schnatter+Malsiner-Walli+Gruen:2020} also discuss this conditional prior for the dynamic MFM.
For the DPM, this prior is independent of $\alpha$:
\begin{linenomath*}
\begin{align*}
  p (N_1, \ldots, N_k |N, K_+ = k) &= \frac{1}{ C ^\infty _{N,k}}
\prod_{j=1}^{k}   \frac{1}{ N_j}.
\end{align*}
\end{linenomath*}
For the static MFM, this prior depends on $\gamma$, but is independent of $p(K)$:
\begin{linenomath*}
\begin{align*}
  p(N_1, \ldots, N_k |N, K_+ = k, \gamma) &= \frac{1}{ C ^{\gamma} _{N,k}}
\prod_{j=1}^{k}   \frac{ \Gamma(N_j +\gamma)}{ \Gamma(N_j + 1)}.
\end{align*}
\end{linenomath*}
For the dynamic MFM, this prior  depends on $\alpha$ as well as on the prior $p(K)$:
\begin{linenomath*}
\begin{align*}
  p (N_1, \ldots, N_k |N, K_+ = k, \alpha) &=  \sum_{K=k}^\infty
w^{K,\alpha}_{N,k} \prod_{j=1}^{k}
\frac{\Gamma(N_j+\frac{\alpha}{K})}{\Gamma(N_j + 1)} ,
\end{align*}
\end{linenomath*}
where
\begin{linenomath*}
\begin{align}  \label{wDyn}
  w^{K,\alpha}_{N,k} &= \frac{\commentR{\check{w}}^{K,\alpha}_{N,k}}{\sum_{K=k}^\infty
\commentR{\check{w}}^{K,\alpha}_{N,k} C ^{K, \alpha} _{N,k}}, &
  \commentR{\check{w}}^{K,\alpha}_{N,k} &=  \frac{ p(K)  K !}{(K- k )!  K^k \Gamma(1 +
\frac{\alpha}{K})^k}.
\end{align}
\end{linenomath*}  
These results suggest that the dynamic MFM has an increased flexibility with respect to the prior on the partitions compared to the static MFM and the DPM. Empirical differences to the dynamic MFM when varying the prior on $K$ are investigated in Section~\ref{sec:insp-induc-priors}.

\subsubsection{Marginalising the prior on the labelled data cluster
sizes}\label{sec:marg-prior-labell}
The marginal \commentBG{conditional} density $P(N_j = n |N, K_+ = k , \bm{\gamma})$ is the same for all $j=1,\ldots,k$. In the following, we obtain without loss of generality $P(N_k = n |N, K_+ = k , \bm{\gamma})$ from $p(N_1, \ldots, N_{k} |N , \bm{\gamma} )$, by summing over all partitions where the size of data cluster $k$ is equal to $n$, i.e., $N_{k} = n$, with $n= 1,\ldots,N-k+1$ and the remaining data cluster sizes sum up to $N-n$, i.e., $N_1+\ldots+N_{k-1}=N-n$:
\begin{linenomath*}
\begin{multline*}
P(N_k = n |N, K_+ = k , \bm{\gamma})  = \frac{P(N_k= n |N, \bm{\gamma}) }{P(K_+ = k |N, \bm{\gamma})}\\
  \displaystyle
  \frac{N!}{ k! P(K_+ = k |N, \bm{\gamma}) }  \sum_{K=k}^\infty   p(K)  \frac{
{V}_{N, k}^{K, \commentF{\gamma_K}}}{ \Gamma(\commentF{\gamma_K}) ^{k} }
\frac{ \Gamma(n + \commentF{\gamma_K} )} {\Gamma(n + 1) }  \sum_{\substack{N_1,\ldots,
N_{k-1}>0 \\N_1+\ldots+N_{k-1}=N-n}}
\prod_{j=1}^{k-1} \frac{ \Gamma(N_{j} + \commentF{\gamma_K} )} {\Gamma(N_j + 1)}.
\end{multline*}
\end{linenomath*}
Using the definition of $C^{K, \gamma_K}_{N,k}$ in Equation~\eqref{PposCk}, we obtain for $n=1,\ldots,N-k+1$:
\begin{linenomath*}
\begin{align*}
  P(N_k= n |N, K_+ = k , \bm{\gamma} )  &=
 \displaystyle \frac{  \sum_{K=k}^\infty   p(K)  \frac{ {V}_{N, k}^{K, \gamma_K}}{ \Gamma(
\gamma_K) ^{k} }
  \frac{ \Gamma(n +\gamma_K)} {\Gamma(n + 1)}  C^{K, \gamma_K}_{N-n,k-1}}
  { \sum_{K=k}^\infty   p(K)  \frac{ {V}_{N, k}^{K, \gamma_K}}{ \Gamma(
\gamma_K ) ^{k}} C^{K, \gamma_K}_{N,k} }.
\end{align*}
\end{linenomath*}
Therefore, the marginal prior can be expressed for $n=1, \ldots, N- k+1$ and $j=1,\ldots,k$ as,
\begin{linenomath*}
\begin{align}  \label{PNj}
P(N_j= n |N, K_+ = k, \bm{\gamma})  &= \sum_{K=k}^\infty   w^{K,\gamma_K}_{N,k}
                                      \frac{ \Gamma(n +\gamma_K)} {\Gamma(n +
1)}  C^{K, \gamma_K}_{N-n,k-1},
\end{align}
\end{linenomath*}
where
\begin{linenomath*}
\begin{align*}
  w^{K,\gamma_K}_{N,k} &= \frac{\tilde{w}^{K,\gamma_K}_{N,k}}{\sum_{K=k}^\infty
 \tilde{w}^{K,\gamma_K}_{N,k} C^{K, \gamma_K}_{N,k}  },\\
\tilde{w}^{K,\gamma_K}_{N,k}&= \frac{p(K) {V}_{N, k}^{K, \gamma_K}}{ \Gamma(
\gamma_K ) ^{k} } =
 \frac{   p(K)   (\gamma_K) ^{k}  \Gamma(\gamma_K K) K !  }{ \Gamma(1+\gamma_K)
^k \Gamma(\gamma_K K+N) (K- k )!}.
\end{align*}
\end{linenomath*}
For the DPM, this simplifies to
\begin{linenomath*}
\begin{align*}
 P(N_j = n |N, K_+ = k) &=
  \frac{1}{n C ^\infty _{N,k}}
 \sum_{\substack{N_1,\ldots, N_{k-1}>0 \\N_1+\ldots+N_{k-1}=N -n}}
\prod_{j=1}^{k-1}   \frac{ 1}{ N_j } =
  \frac{C ^\infty_{N-n,k-1}}{n C ^\infty _{N,k}}.
\end{align*}
\end{linenomath*}
For the static MFM, this prior is given by
\begin{linenomath*}
\begin{align*}
  P(N_j = n |N, K_+ = k, \gamma) & =
  \frac{ \Gamma(n +\gamma)}{ \Gamma(n + 1) C ^{\gamma} _{N,k}}
  \sum_{\substack{N_1,\ldots, N_{k-1} > 0\\N_1+\ldots+N_{k-1}=N -n}}
\prod_{j=1}^{k-1}   \frac{ \Gamma(N_j +\gamma)}{ \Gamma(N_j + 1)} \\
  & =  \frac{ \Gamma(n +\gamma)}{ \Gamma(n + 1)} \frac{C ^{\gamma} _{N-n,k-1}}{
C ^{\gamma} _{N,k}} .
\end{align*}
\end{linenomath*}
For the dynamic MFM, this is equal to
\begin{linenomath*}
\begin{align*}
   P(N_j = n |N, K_+ = k, \alpha) &= \sum_{K=k}^\infty
w^{K,\alpha}_{N,k}\frac{\Gamma(n+\frac{\alpha}{K})}{\Gamma(n + 1)} C ^{K,
\alpha} _{N-n,k-1} ,
\end{align*}
\end{linenomath*}
 \commentR{where $w^{K,\alpha}_{N,k}$ is the same as in (\ref{wDyn}).}\footnote{\commentR{Note that $\tilde{w}^{K,\alpha}_{N,k}= \frac{\alpha ^k \Gamma (\alpha) }{\Gamma (\alpha+N) } \check{w}^{K,\alpha}_{N,k}$ and the first factor cancels when normalising $ \tilde{w}^{K,\alpha}_{N,k}$ to obtain $w^{K,\alpha}_{N,k}$.}}
 Compared to the prior on the number of data clusters $K_+$, this implies that for the dynamic MFM, for each specific number of data clusters $k$, $C^{K, \gamma_K} _{N-n,k-1}$ does not only need to be determined depending on $K$, but also for $N-n$ with $n=1,\ldots,N-k+1$. For $w^{K,\gamma_K}_{N,k}$, $C^{K, \gamma_K} _{N,k}$ also needs to be determined. In the case of the static MFM and the DPM, the computation is less involved as \commentF{$C^{K, \gamma_K} _{\tilde{n},k-1}, \tilde{n}=k-1, \ldots, N-1$,} and $C^{K, \gamma_K} _{N,k}$ do not depend on $K$.

\subsubsection{Computing conditional prior means for function\commentBG{s of a single or two} data cluster sizes}\label{sec:comp-prior-moments}

The computation of the prior expectation $\mathbb{E}(\psi(N_j)|N, K_+ = k, \bm{\gamma})$ of any function $\psi(N_j)$ with respect to the conditional prior on the labelled data cluster sizes is straightforward, given the marginal prior $P(N_j = n |N, K_+ = k, \bm{\gamma})$ derived in Equation~\eqref{PNj}:
\begin{linenomath*}
\begin{align}
  \mathbb{E}(\psi(N_j)|N, K_+ = k, \bm{\gamma}) &=  \sum_{n=1} ^{N- k+1}
\psi(n) P(N_j = n |N, K_+ = k, \bm{\gamma})  \nonumber\\
  & =  \sum_{K=k}^\infty w^{K,\gamma_K}_{N,k} \sum_{n=1} ^{N- k+1} \psi(n)
\frac{ \Gamma(n +\gamma_K)}{ \Gamma(n + 1)}C ^{K, \gamma_K} _{N-n,k-1} .
\label{EWj}
\end{align}
\end{linenomath*}
Note that $\mathbb{E}(\psi(N_j)|N, K_+ = k, \bm{\gamma})$ is the same for all $j=1, \ldots, k$.

The sequence $C^{K, \gamma_K}_{N- n,k-1}, n=1, \ldots, N- k+1 $ results for each $K$ as a byproduct of recursion \eqref{recck} in Algorithm~\ref{KNMFM} in Appendix~\ref{sec:algor-comp-prior}, since
\begin{linenomath*}
\begin{align*}
  \bm{c} _{K,k-1}   &=   \left( C_{N,k-1}  ^{K, \gamma_K}, C_{N-1,k-1}   ^{K,
\gamma_K}, C_{N-2,k-2}   ^{K, \gamma_K}, \ldots,   C_{k-1,k-1}  ^{K, \gamma_K}
\right)^{\top}.
\end{align*}
\end{linenomath*}
Hence, the recursion in Algorithm~\ref{KNMFM} in Appendix~\ref{sec:algor-comp-prior} can be applied for each $K$ to determine $\bm{c} _{K,k-1} $.
Removing the first element of $\bm{c} _{K,k-1} $ yields then the $(N-k+1)$-dimensional vector $\tilde{\bm{c}}_{K,k-1} = (C^{K, \gamma_K}_{N-1,k-1},\ldots, C^{K, \gamma_K}_{k-1,k-1} )^{\top} $. $\mathbb{E}(\psi(N_1) |N, K_+ = k, \bm{\gamma})$ is thus computed efficiently using:
\begin{linenomath*}
\begin{align}\label{eq:algorithm1}
\mathbb{E}(\psi(N_1) |N, K_+ = k, \bm{\gamma})
&= \sum_{K=k}^\infty w^{K,\gamma_K}_{N,k} \tilde{\bm{c}}_{K,k-1}^{\top}
\bm{a}_k,
\end{align}
\end{linenomath*}
where $\bm{a} _k$ is an $(N-k+1)$-dimensional vector defined in Equation~\eqref{eq:a_k} with $ a_n = \tilde \psi ( n) $ and
\begin{linenomath*}
\begin{align}\label{eq:tildepsi}
  \tilde\psi (x) = \frac{\psi(x) \Gamma(x+\gamma_K)}{\Gamma(x + 1)}.
\end{align}
\end{linenomath*}
Next, we investigate how to determine \commentBG{the expectation} $\mathbb{E}(\psi(N_j)\psi(N_\ell) |N, K_+ = k, \bm{\gamma}) $ for $j\neq \ell$. For $k=2$, we can use that $N_{2} =N-N_1$, hence
\begin{linenomath*}
\begin{align*}
\psi (N_1) \psi (N_2) = N_1(\log N_1)  N_{2}  (\log N_{2})= N_1 (N- N_1) \log  N_1 \, \log (N-N_1)
\end{align*}
\end{linenomath*}
depends only on $N_1$ and Equation~\eqref{EWj} can be used to compute $\mathbb{E}(\psi(N_1)\psi(N_2) |N, K_+ = 2, \bm{\gamma})$.

For $k \geq 3$, the bivariate marginal prior $p(N_1, N_2|N, K_+ = k, \bm{\gamma})$ is given for all pairs $\{(N_1,N_2): 2 \leq N_1+N_2 \leq N-k+2)\}$ by:
\begin{linenomath*}
\begin{align*}
p(N_1, N_2|N, K_+ = k, \bm{\gamma})
&= \sum_{K=k}^\infty w^{K,\gamma_K}_{N,k} \left[\prod_{j=1}^{2}
\frac{\Gamma(N_j+\gamma_K)}{\Gamma(N_j + 1)}\right]
C_{N-N_1-N_2,k-2}^{K, \gamma_K},
\end{align*}
\end{linenomath*}
where $w^{K,\gamma_K}_{N,k}$ are the same weights as in Equation~\eqref{PNj}. In principle, $\mathbb{E}(\psi(N_1)\psi(N_{2}) |N, K_+ = k, \bm{\gamma})$ is obtained by summing $p(N_1, N_2|N, K_+ = k, \bm{\gamma})$ over all possible pairs $(N_1,N_2)$:
\begin{linenomath*}
\begin{multline*}
\mathbb{E}(\psi(N_1)\psi(N_{2}) |N, K_+ = k, \bm{\gamma})
=\\
\sum_{K=k}^\infty w^{K,\gamma_K}_{N,k} \sum_{n_1=1} ^{N- k+1}
\sum_{n_2=1} ^{N- n_1 - k+2} \prod_{j=1}^{2} \frac{\psi(n_j)
\Gamma(n_j+\gamma_K)}{\Gamma(n_j + 1)}
C_{N-n_1-n_2,k-2}^{K, \gamma_K}.
\end{multline*}
\end{linenomath*}
It is convenient to arrange the enumeration such that one sums over
$n =n_1+n_2$:
\begin{linenomath*}
\begin{align*}
\mathbb{E}(\psi(N_1)\psi(N_{2}) |N, K_+ = k, \bm{\gamma})
&= \sum_{K=k}^\infty w^{K,\gamma_K}_{N,k}
\sum_{ n =2} ^{N- k+2} C^{K, \gamma_K}_{N-  n,k-2}  \sum_{ m =1} ^{ n -1}
\tilde \psi ( m) \tilde\psi (n -  m),
\end{align*}
\end{linenomath*}
where again $\tilde\psi (x)$ is as defined in Equation~\eqref{eq:tildepsi}.

The sequence of inner sums $\sum_{ m =1} ^{ n -1} \tilde \psi ( m) \tilde\psi (n - m)$ for $n=2,\ldots,N-k+2$ corresponds to the vector resulting from multiplying the matrix $\bm{A}_k$ with the vector $\bm{a}_k$ where $\bm{A} _k$ is a $(N-k+1) \times (N-k+1)$ lower triangular Toeplitz matrix and $\bm{a} _k$ is the $(N-k+1)$-dimensional vector defined as
\begin{linenomath*}
\begin{align}
\bm{A}_k  &= \left(
\begin{array}{lllll}
a_1 &   &   & &  \\
a_2 & a_1      &   &               &  \\
\vdots   &     \ddots    & \ddots  &                 &   \\
a_{N-k}   &               &   a_2        &  a_1               &  \\
a_{N-k+1} &      \ddots          &      \ddots      &       a_2
&  a_1 \\
\end{array}
\right), \qquad
  \bm{a}_k  = \left( \begin{array}{c} a_1 \\ \vdots \\ a_{N-k+1}  \end{array}
\right),
  \label{eq:a_k}
\end{align}
\end{linenomath*}
where $ a_n = \tilde \psi ( n) $.  The sequence $C^{K, \gamma_K}_{N- n,k-2}, n=2, \ldots, N- k+2 $ results for each $K$ as a byproduct of recursion \eqref{recck} in Algorithm~\ref{KNMFM} in Appendix~\ref{sec:algor-comp-prior}, since
\begin{linenomath*}
\begin{align*}
\bm{c} _{K,k-2}   &=   \left(C_{N,k-2}  ^{K, \gamma_K},  C_{N-1,k-2}   ^{K,
\gamma_K}, C_{N-2,k-2}   ^{K, \gamma_K}, \ldots,  C_{k-2,k-2}  ^{K,
\gamma_K}\right)^{\top}.
\end{align*}
\end{linenomath*}
Hence, the recursion in Algorithm~\ref{KNMFM} in Appendix~\ref{sec:algor-comp-prior} is applied for each $K$ to determine $\bm{c} _{K,k-2} $.
Removing the first two elements of $\bm{c} _{K,k-2} $ yields then the $(N-k+1)$-dimensional vector $\check{\bm{c}}_{K,k-2} = (C^{K, \gamma_K}_{N- 2,k-2},\ldots, C^{K,\gamma_K}_{k-2,k-2} )^{\top} $.
$\mathbb{E}(\psi(N_1)\psi(N_{2}) |N, K_+ = k, \bm{\gamma})$ is computed efficiently using:
\begin{linenomath*}
\begin{align}\label{eq:algorithm2}
\mathbb{E}(\psi(N_1)\psi(N_{2}) |N, K_+ = k, \bm{\gamma})
&= \sum_{K=k}^\infty w^{K,\gamma_K}_{N,k} \check{\bm{c}}_{K,k-2}^{\top}
\bm{A}_k  \bm{a}_k.
\end{align}
\end{linenomath*}
Again $\mathbb{E}(\psi(N_j)\psi(N_{\ell}) |N, K_+ = k, \bm{\gamma})$, is the same for all $j,\ell=1,\ldots,k$, $j \neq \ell$ and thus given
by Equation~\eqref{eq:algorithm2}.

\subsubsection{Computing the prior mean and variance of the functionals conditional on $K_+$}\label{sec:computing-prior-mean}

With Equations \eqref{eq:algorithm1} and \eqref{eq:algorithm2}, the first two moments of $\Psi(N_1,\ldots,N_k)$ conditional on $K_+ = k$ can be calculated efficiently. In the following, we derive the conditional mean and variance of $\Psi$ written in terms of quantities derived in Section~\ref{sec:comp-prior-moments}. Additionally, two empirically relevant examples \commentBG{for functionals} $\Psi$ \commentBG{which allow to characterise} the prior on the partitions are introduced. One of the examples is the relative entropy suggested by \citet{Green+Richardson:2001} and the other is the number of singletons in the partitions. In Sections~\ref{sec:insp-induc-priors} and \ref{sec:comp-defa-priors}, these two functionals are evaluated for all three \commentBG{Bayesian mixture} models \commentBG{with various prior} settings.

The prior mean and variance of $\Psi(N_1,\ldots,N_k)$ conditional on $K_+ = k$ as well as $N$ and $\bm{\gamma}$ are given by
\begin{linenomath*}
\begin{align}  \label{EFUN}
  \mathbb{E}(\Psi(N_1, \ldots,N_k)|N, K_+ = k, \bm{\gamma})
   &= k \mathbb{E}(\psi(N_j)|N, K_+ = k ,  \bm{\gamma}),\\
 \label{VFUN}
  \mathbb{V}(\Psi(N_1, \ldots,N_k)|N, K_+ = k,  \bm{\gamma})
  &=
k \mathbb{E}(\psi(N_j)^2|N, K_+ = k , \bm{\gamma}) + \\ \nonumber
 k(k-1) \mathbb{E}(\psi(N_j)  \psi(N_{\ell})&|N, K_+ = k , \bm{\gamma})
  -
k^2 (\mathbb{E}(\psi(N_j)|N, K_+ = k , \bm{\gamma}))^2,
\end{align}
\end{linenomath*}
with $j\neq \ell$.

The expectation in Equation~\eqref{EFUN} and all expectations in Equation~\eqref{VFUN} involving a single data cluster size $N_j$ are evaluated efficiently with Equation~\eqref{eq:algorithm1} while $\mathbb{E}(\psi(N_j) \psi(N_{\ell}) |N, K_+ = k, \bm{\gamma})$ is computed using Equation~\eqref{eq:algorithm2}.

%
\paragraph{Relative entropy.}
The relative entropy in a partition with a fixed number $k$ of data clusters is defined as
\begin{linenomath*}
\begin{align*}
\mathcal{E}(N_1, \ldots, N_k) / \log k = - \frac{1}{\log k } \sum_{j=1}^k
\frac{N_j}{N} \log \frac{N_j}{N} = -\frac{1}{N \log k} \sum_{j=1}^k N_j  \log
N_j + \frac{\log N}{\log k}.
\end{align*}
\end{linenomath*}
Regardless of $k$, the relative entropy takes values in (0, 1] with values close to 1 indicating similarly large data cluster sizes $N_1,\ldots, N_k$.  For the most balanced clustering where all $N_j$, $j=1,\ldots,k$ are equal, the relative entropy is exactly equal to 1.
Higher prior mean values indicate that a-priori more balanced partitions are induced, while larger prior variance or standard deviation values indicate that the prior partition distribution is more flexible.

The calculation of the relative entropy is based on the functional $\psi(N_j)=N_j \log N_j$. The prior expectation of the relative entropy is equal to $\mathbb{E}_{\mathcal{E},k} = \mathbb{E}(\mathcal{E}(N_1, \ldots,N_k)|N, K_+ =  k, \bm{\gamma})/\log k $ with
\begin{linenomath*}
\begin{align*}
  \mathbb{E}(\mathcal{E}(N_1, \ldots,N_k)|N, K_+ = k, \bm{\gamma})
                           &= \log N - \frac{k}{N} \mathbb{E}(N_j \log N_j|N,
K_+ = k ,  \bm{\gamma}).
\end{align*}
\end{linenomath*}
The prior variance of the relative entropy is equal to $\mathbb{V}_{\mathcal{E},k} = \mathbb{V}(\mathcal{E}(N_1, \ldots,N_k)|N, K_+ = k, \bm{\gamma})/(\log k)^2$ with
\begin{linenomath*}
\begin{multline*}
  \mathbb{V}(\mathcal{E}(N_1, \ldots,N_k)|N, K_+ = k,  \bm{\gamma})
  =
  \frac{1}{N^2}\left(
k \mathbb{E}(N_j^2 (\log N_j)^2|N, K_+ = k , \bm{\gamma})
+  \right. \\
 k(k-1) \mathbb{E}(N_j(\log N_j)  N_{\ell}  (\log N_{\ell}) |N, K_+ = k ,
\bm{\gamma})
 -  \\
 \left.
 k^2 (\mathbb{E}(N_j \log N_j|N, K_+ = k , \bm{\gamma}))^2  \right),
\end{multline*}
\end{linenomath*}
where $j\neq \ell$.
\paragraph{Number of singletons.}
The calculation of the number of singletons is based on the functional $\psi(N_j)= \vmathbb{1}_{\{N_j = 1\}}$, where $\vmathbb{1}$ is the indicator function. The prior mean and variance are straightforward to calculate by plugging the functional into Equations~\eqref{EFUN} and \eqref{VFUN}.

\subsubsection{Computing the prior mean and variance of the
functionals unconditional on $K_+$}\label{sec:computing-prior-mean-uncond}

With the distribution of $p(K_+|N,\bm{\gamma})$ derived in Section~\ref{sec:induced-prior-number} coupled with the first two moments of symmetric additive functionals conditional on $K_+$ shown in Section~\ref{sec:computing-prior-mean}, we are now ready to marginalise out $K_+$ to obtain the first two moments of symmetric additive functionals unconditional on $K_+$.

By the law of total expectation, the prior mean of $\Psi(N_1,\ldots,N_{K_+})$ unconditional on $K_+$ can be computed trivially combining $\mathbb{E}_k \equiv \mathbb{E}(\Psi(N_1,\ldots,N_k)|N,K_+ = k,\bm{\gamma})$ derived in Equation \eqref{EFUN} and $P(K_+ = k|N,\bm{\gamma})$ derived in Equation \eqref{pKplus} as follows:
\begin{linenomath*}
\begin{align}
    \mathbb{E}(\Psi(N_1,\ldots,N_{K_+})|N,\bm{\gamma}) = \sum_{k = 1}^N \mathbb{E}_k\label{EFUNuncond}P(K_+ = k|N,\bm{\gamma}).
\end{align}
\end{linenomath*}
Following the same line of reasoning, by the law of total variance, the prior variance of $\Psi(N_1,\ldots,N_{K_+})$ unconditional on $K_+$ can be obtained from Equations \eqref{VFUN} and \eqref{pKplus} as follows:
\begin{linenomath*}
\begin{multline}
    \mathbb{V}(\Psi(N_1,\ldots,N_{K_+})|N,\bm{\gamma}) = \sum_{k = 1}^{N-1}\mathbb{V}_kP(K_+ = k|N,\bm{\gamma})+\label{VFUNuncond}\\
    \sum_{k = 1}^N \mathbb{E}_kP(K_+ = k |N, \bm{\gamma})(1-P(K_+ = k |N, \bm{\gamma}))-\\
    2\sum_{k = 2}^{N-1}\sum_{k'=1}^{k-1}\mathbb{E}_k\mathbb{E}_{k'}P(K_+ = k |N, \bm{\gamma})P(K_+ = k' |N, \bm{\gamma}),
\end{multline}
\end{linenomath*}
where $\mathbb{V}_k \equiv \mathbb{V}(\Psi(N_1,\ldots,N_k)|N,K_+ = k,\bm{\gamma})$ and the sum is taken over $k = 1$ to $N-1$ in the variance term of Equation \eqref{VFUNuncond} since the variance of any functional  is trivially 0 \commentBG{for $k = N$, because there exists only one partition which separates $N$ data points into $N$ groups}.

For example, if one wants to compute the prior mean and variance of the relative entropy unconditional on $K_+ =k$, $\mathbb{E}_k$ can be set to $\mathbb{E}_{\mathcal{E},k}/\log(k)$ introduced in the previous section and similarly $\mathbb{V}_k$ \commentBG{can be set} to $\mathbb{V}_{\mathcal{E},k}/(\log k)^2$.

\section{Empirical inspection of the induced priors}\label{sec:insp-induc-priors}

This section serves as a demonstration of the proposed methodology where the static and dynamic MFMs with respect to various aspects of their induced prior on the partitions are compared utilising all the tools introduced in Section~\ref{sec:induced-priors}. The DPM is not included in this comparison as it assumes infinitely many clusters in the population as opposed to MFM models with \commentBG{a} finite but unknown number of clusters $K$ \commentBG{in the population}. The comparison specifically involves the prior on the number of data clusters $K_+$ and the prior moments of several functionals computed over the partitions for these two MFM models. These implicit finite-sample characteristics are induced by three \commentBG{data and} model specifications: the sample size $N$, the prior on the number of clusters in the population $p(K)$ and the hyperparameter of the prior component weight distribution $\gamma$ or $\alpha$ depending on the type of the MFM.

In the following, we systematically compare these induced priors by considering several different priors on $K$: a uniform distribution on $[1, 30]$ \citep{Richardson+Green:1997} for $K$, the geometric distribution \commentF{$\mbox{\rm Geo}(0.1)$} 
 for $K-1$ \citep{Miller+Harrison:2018} and the beta-negative-binomial distribution BNB$(1, 4, 3)$ for $K-1$ \citep{Fruehwirth-Schnatter+Malsiner-Walli+Gruen:2020}.
%
The sample size is fixed to $N = 100$ across all comparisons considered in this section.

\subsection{Comparing the prior on the number of data clusters
$K_+$}\label{sec:revis-prior-numb}

\begin{figure}[t!]
  \centering
  \includegraphics[width=\textwidth]{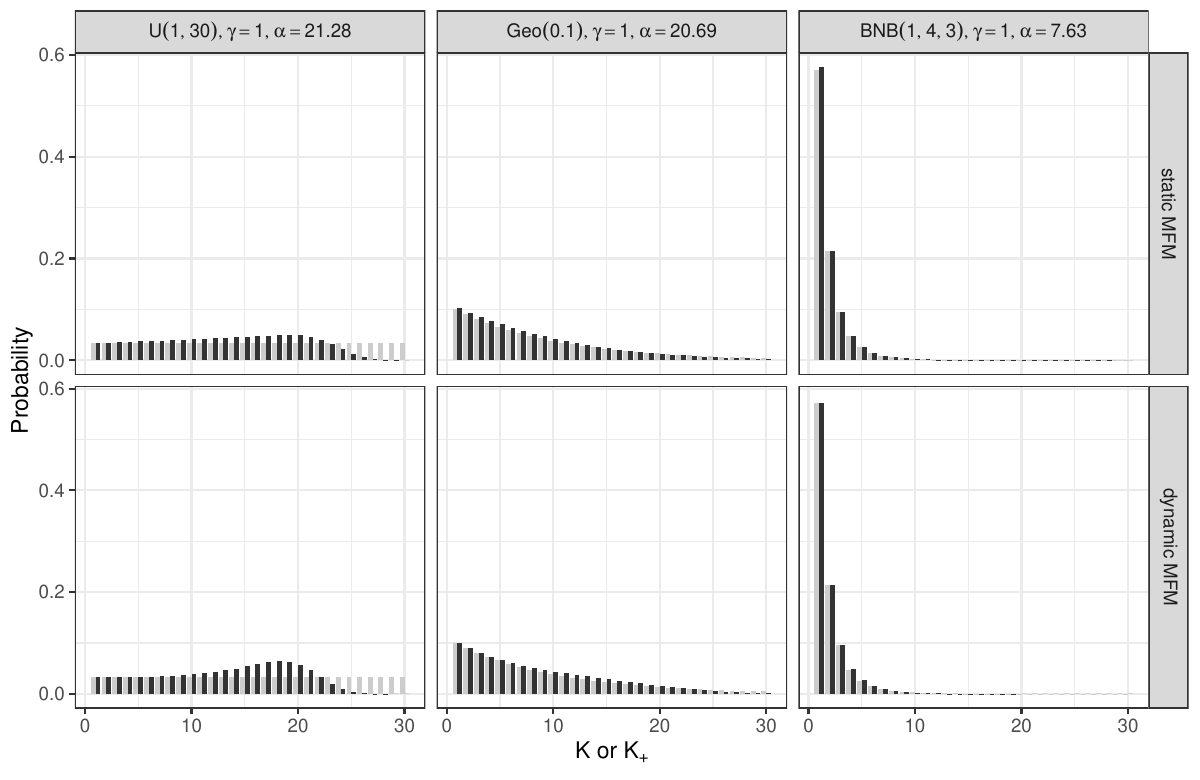}
  \caption{\label{fig:figure1a-moment-rev}The prior probabilities of
    $K$ (in grey) and $K_+$ (in black) for different priors on $K$ and
    $N=100$, for the static MFM with $\gamma = 1$ and the dynamic
    MFM where $\alpha$ is specified to induce the same prior mean value
    for $K_+$.}
\end{figure}

\begin{figure}[t!]
  \centering
  \includegraphics[width=\textwidth]{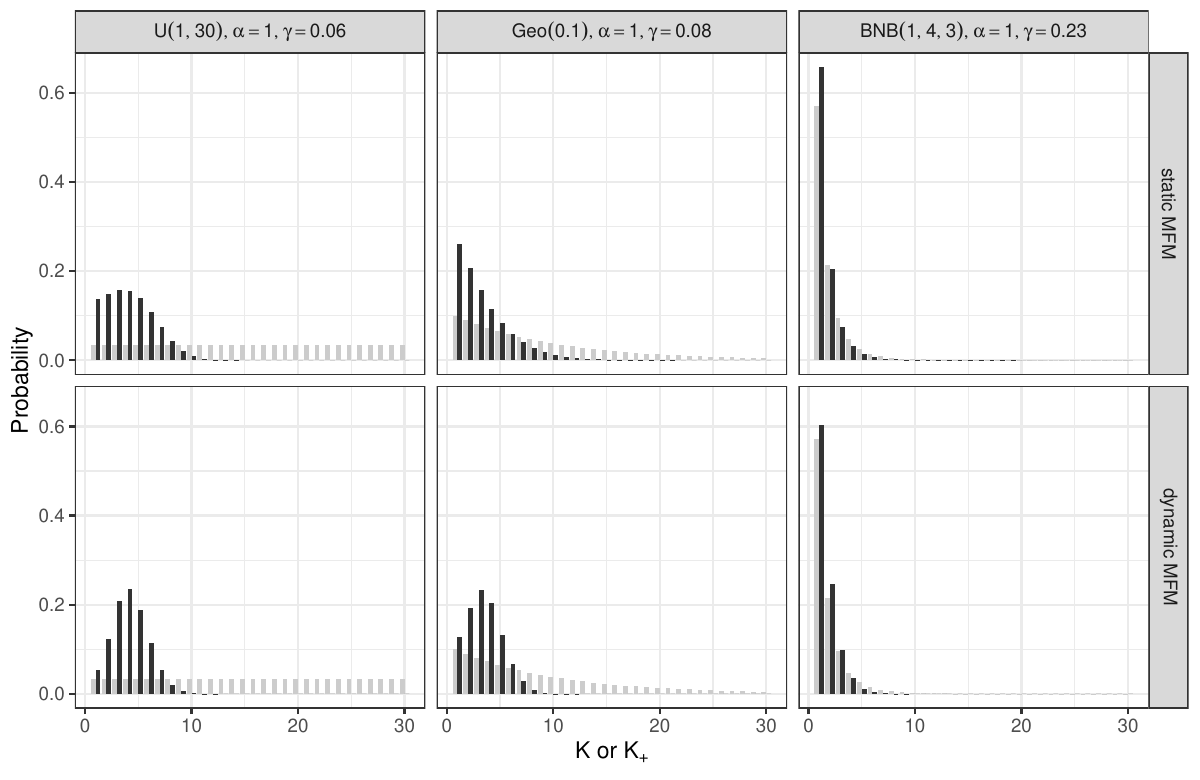}
  \caption{\label{fig:figure1b-moment-rev}The prior probabilities of
    $K$ (in grey) and $K_+$ (in black) for different priors on $K$ and
    $N=100$, for the dynamic MFM with $\alpha = 1$ and the static MFM
    where $\gamma$ is specified to induce the same prior mean value for
    $K_+$.}
\end{figure}

To begin with, the prior distribution on $K_+$ is compared between the static and dynamic MFMs. However, when comparing the prior on $K_+$ between the two MFM models, the incomparability of the Dirichlet parameter $\gamma$ and $\alpha$ obstructs a direct comparison if those are fixed to the same value. For this reason, moment matching with respect to their prior mean of $K_+$ is done twice: \commentF{on  the one hand,} $\alpha$ of the dynamic MFM is chosen to match the static MFM with $\gamma = 1$, \commentF{and on the other hand,}  $\gamma$ of the static MFM is matched \commentBG{to} the dynamic MFM with $\alpha = 1$. The former results are presented in Figure~\ref{fig:figure1a-moment-rev} and the latter in Figure~\ref{fig:figure1b-moment-rev}.

Figure~\ref{fig:figure1a-moment-rev} indicates that no perceivable differences in the distribution of $K_+$ (black bars) can be seen between the static MFM in the top row and the dynamic MFM in the bottom row \commentBG{within the same column, where each columns represents} a different prior on $K$. Under this setting, the prior on $K_+$ traces the prior on $K$ to a certain extent for all cases. 
On the other hand, in Figure~\ref{fig:figure1b-moment-rev}, the static and dynamic MFM models differ greatly in \commentBG{their} priors on $K_+$ when the uniform and geometric prior are assigned on $K$. Noticeably, for the uniform and geometric prior cases, the dynamic MFM assigns less mass to the probability of homogeneity, that is, $p(K_+ = 1|N,\bm{\gamma})$. \commentBG{Also for the uniform and geometric prior on $K$, a clear difference between the priors on $K$ and $K_+$ is visibly with a lot more mass assigned to small values of $K_+$ than of $K$.}

To summarise, for both MFM models, the characteristics of the prior on $K_+$ are sometimes markedly different from that of $K$. Especially, those of the dynamic MFM departs quite considerably from the prior on $K$ under certain combinations of $\alpha$ and $p(K)$ as demonstrated in Figure~\ref{fig:figure1b-moment-rev}. \commentBG{This} strongly indicates the need of using the proposed methodology to investigate the induced prior on $K_+$.

\subsection{Comparing the prior on the partitions based on
    symmetric additive functionals}\label{sec:revis-prior-part}

Section~\ref{sec:char-prior-part} introduced procedures to compute the prior mean and variance of any symmetric additive functionals over the induced prior partitions. Here, we specifically consider two functionals introduced there: the relative entropy and the number of singletons in the partitions.
\begin{figure}[h!]
  \centering
  \includegraphics[width=\textwidth]{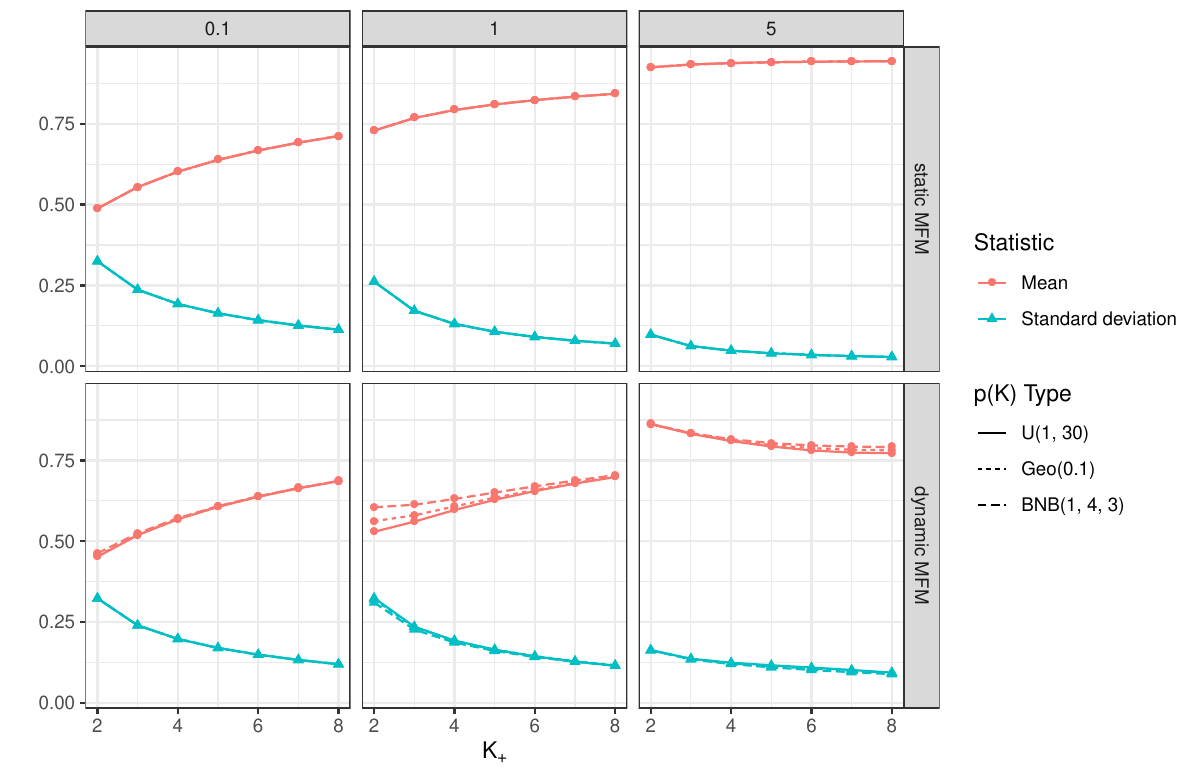}
  \caption{\label{fig:entropy-a-rev} The prior mean and standard
    deviation of the relative entropy of the partitions for
    $K_+ \in [2, 8]$, for three different
    priors on $K$, for the static and dynamic MFM with $\gamma$ or
    $\alpha \in \{0.1, 1, 5\}$ (from left to right) and $N = 100$.}
\end{figure}

Each plot of Figure~\ref{fig:entropy-a-rev} shows the prior mean and
standard deviation obtained for the relative entropy of the partition
distribution conditional on a specific number of data clusters $K_+$
ranging from 2 to 8. Column-wise, these figures are arranged in order
of magnitude of their corresponding hyperparameter $\gamma$ or
$\alpha$ increasing from left to right whereas each row represents the
specific MFM model. As shown in Section~\ref{sec:induc-cond-prior-1},
for the static MFM, the prior on $K$ does not have any impact once
conditioned on $K_+$. Therefore, figures in the top row do not vary by
$p(K)$ as opposed to those in the bottom row representing results for
the dynamic MFM which seem to \commentBG{slightly vary} depending on
the prior on $K$, \commentBG{in particular} when $\alpha = 1$.

It can be seen that for \commentBG{the static} MFM model, the prior
mean of the relative entropy increases for larger values of $K_+$ and
also for greater values of $\gamma$ or $\alpha$. Conversely, the prior
standard deviation drops relative to these changes in $K_+$ as well as
$\gamma$ or $\alpha$. \commentBG{Results are similar for the dynamic
  MFM if the Dirichlet parameter is small, i.e., $\alpha = 0.1$. The
  larger $\alpha$ the stronger seems to be the difference to the
  static MFM. For $\alpha = 5$ the prior mean is even decreasing for
  increasing $K_+$.}
\begin{figure}[h!]
  \includegraphics[width=\textwidth]{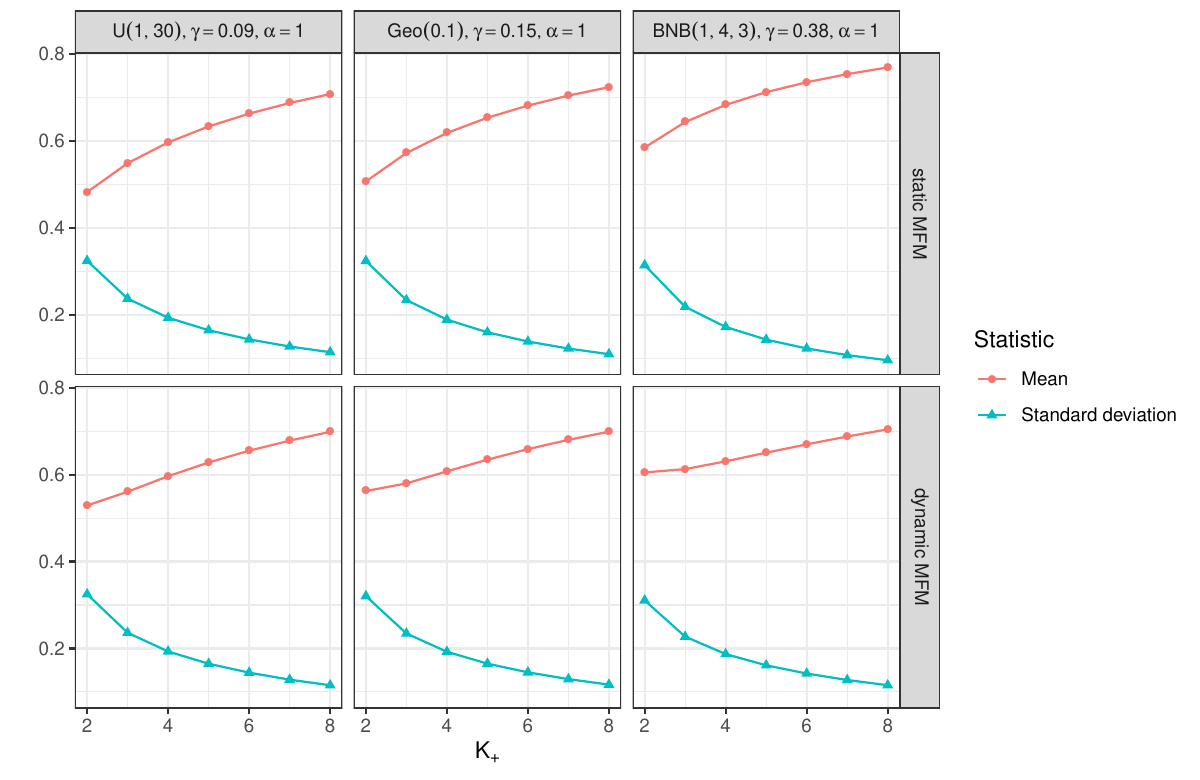}
  \caption{\label{fig:entropy-matched-rev} The prior mean and standard
    deviation of the relative entropy of the partitions for
    $K_+ \in [2, 8]$, for different
    priors on $K$, for the dynamic MFM with $\alpha = 1$ and the corresponding moment matched static MFM, all conditional on $N = 100$.}
\end{figure}

To perform side-by-side comparison between the static and dynamic MFM
models, we focus on the specific setting where the hyperparameter of
the dynamic MFM is fixed to $\alpha = 1$ while considering
\commentBG{again the same} three \commentBG{different} priors on
$K$. The corresponding static MFM for each specification is chosen by
employing the moment matching approach with respect to the
unconditional (with respect to $K_+$) mean of the relative
entropy. Specifically, $\gamma$ of each static MFM is chosen to match
the corresponding dynamic MFM with $\alpha = 1$. By going over the
results summarised in Figure~\ref{fig:entropy-matched-rev}
column-wise, it is clearly \commentBG{visible} that \commentBG{after matching, the
  differences between} each pair of the static MFM and
its dynamic counterpart \commentBG{are negligible} in terms of their prior
means and standard deviations conditional on each $K_+$. 
%
\begin{figure}[h!]
  \includegraphics[width=\textwidth]{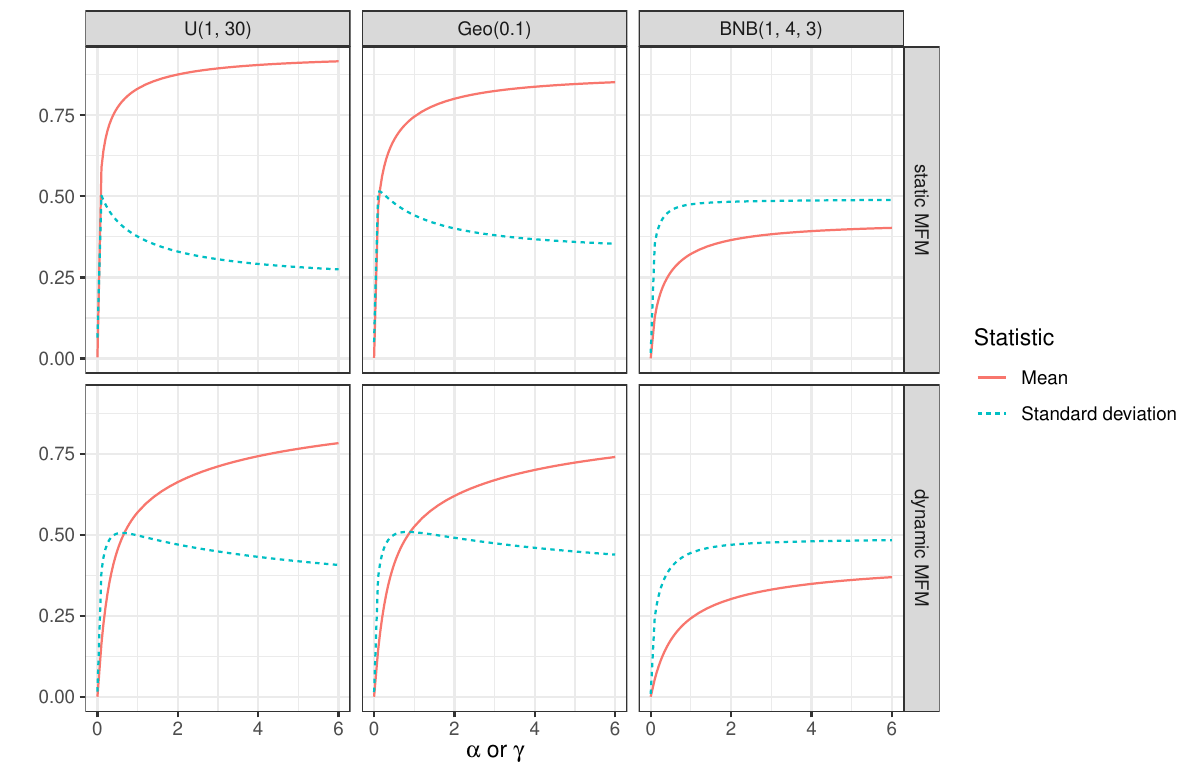}
  \caption{\label{fig:weightedentropy-rev} The prior mean and standard
    deviation of the relative entropy unconditional on $K_+$, computed over the partitions in
    dependence of $\gamma$ or $\alpha$ for different priors on $K$ and
    the static and dynamic MFM, with $N = 100$.}
\end{figure}

To complete the analysis involving the relative entropy, the
functional is computed over the induced prior on the partitions
unconditional on $K_+$ (see
Section~\ref{sec:computing-prior-mean-uncond}). Again, this quantity
is evaluated for the three priors on $K$ in combination with the
static or dynamic MFM for increasing values of $\gamma$ or $\alpha$
\commentBG{and results are} shown in
Figure~\ref{fig:weightedentropy-rev}. For the static MFM, the
conditional relative entropy of the partitions does not vary with
respect to $p(K)$ (see also the theoretic derivations in
Section~\ref{sec:induc-cond-prior-1} and
Figure~\ref{fig:entropy-a-rev}). Therefore, differences in both the
mean and standard deviation among the \commentBG{results for the}
static MFMs shown in \commentBG{the different columns of the top row
  of} Figure~\ref{fig:weightedentropy-rev} originate from differences
in the induced prior on $K_+$. One can clearly observe that these
specifications do imply partitions \commentBG{with rather similar
  characteristics for the uniform and the geometric prior on $K$,
  whereas in particular the prior mean relative entropy is much lower
  for the BNB prior on $K$. In particular for the uniform and
  geometric prior on $K$ the standard deviation peaks for a rather
  small value of $\gamma$ with a sharp decrease followed by a
  levelling off.}  For the dynamic MFM, not only the induced prior on
$K_+$, but also the conditional relative entropy of the partitions
depend on $p(K)$. Thus, one might expect even greater differences in
the prior mean and standard deviation of the relative entropy among
the three dynamic specifications. However, the bottom row results in
Figure~\ref{fig:weightedentropy-rev} suggest that the level of
variability in the prior partition unevenness \commentBG{is to some
  extent comparable for the static and the dynamic MFM}.

\begin{figure}[t!]
  \centering
  \includegraphics[width=\textwidth]{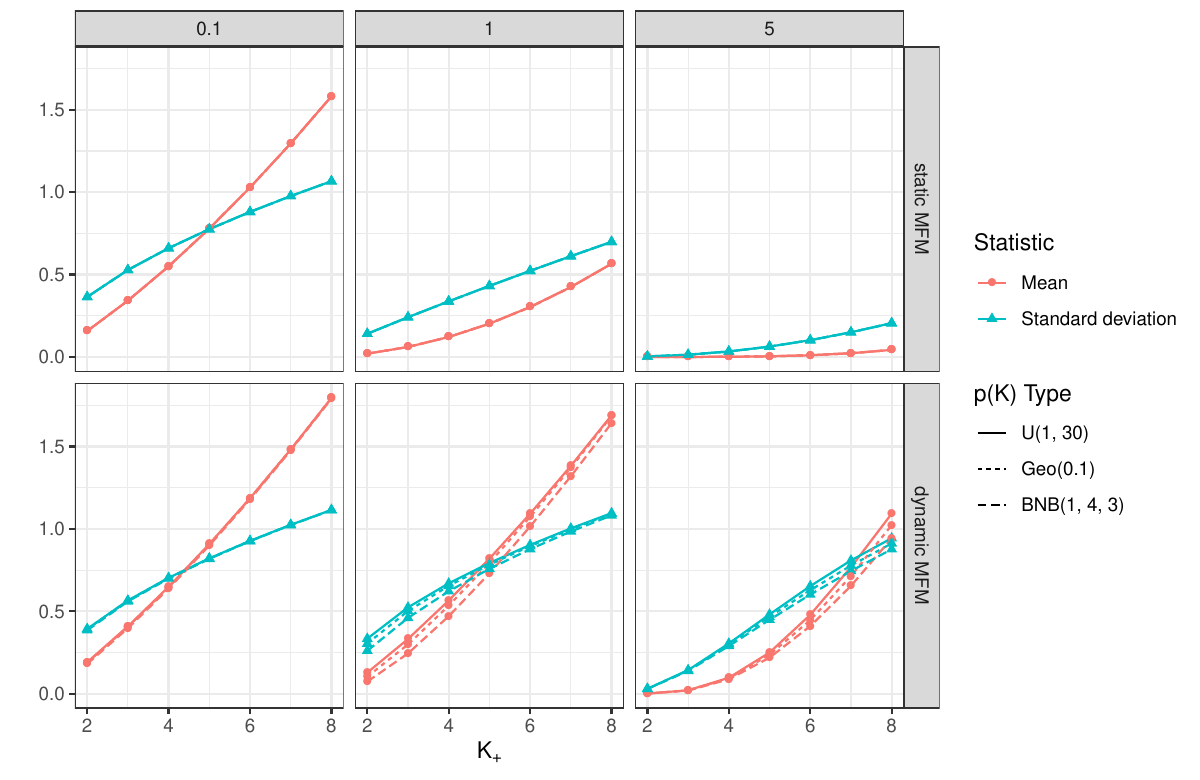}
  \caption{\label{fig:singleton-rev}The prior mean and standard
    deviation of the number of singletons, for different priors on $K$
    and the static and dynamic MFM, with $\gamma$ or
    $\alpha \in \{0.1, 1, 5\}$ and $N = 100$ for
    $K_+ \in [2, 8]$.}
\end{figure}

Another functional of interest is the number of singletons computed
over the induced prior partitions. The results are shown in
Figure~\ref{fig:singleton-rev} \commentBG{using} the same format as in
Figure~\ref{fig:entropy-a-rev}, now with \commentF{the} $y$-axis
representing the prior mean and standard deviation of the number of
singleton clusters. Again, the results for the static MFM do not
depend on $p(K)$. Those of the dynamic MFM depend on $p(K)$ but only
slightly, with the largest impact of $p(K)$ being again observable for
$\alpha = 1$. For both MFMs, an increase in the component weight
hyperparameter $\gamma$ or $\alpha$ corresponds to an overall decrease
in the expected number of singletons and its variability. This is
expected as the partition\commentBG{s} will be more evenly balanced as
$\gamma$ or $\alpha$ increases, see
Figure~\ref{fig:entropy-a-rev}. This means that partitions given high
prior probability will mainly \commentBG{contain} clusters with more than one
observation.

\section{Comparing default priors in Bayesian cluster
analysis}\label{sec:comp-defa-priors}
The main focus of this section is to understand the
induced prior on the partitions when defining specific prior
combinations as suggested in the previous literature for the three
models introduced in Section~\ref{sec:explicit-priors}:
\begin{enumerate}
\item DPMs with $\alpha = 1/3$ \citep[see][]{Escobar+West:1995}.
\item Static MFMs with a uniform prior $[1, 30]$ on $K$ and
  $\gamma = 1$ \citep[see][]{Richardson+Green:1997}.
\item Dynamic MFMs with a BNB$(1, 4, 3)$ prior on $K-1$ and
  $\alpha = 2/5$
  \citep[see][]{Fruehwirth-Schnatter+Malsiner-Walli+Gruen:2020}.
\end{enumerate}
Note that \citet{Escobar+West:1995} and \citet{Fruehwirth-Schnatter+Malsiner-Walli+Gruen:2020} propose a hyperprior on $\alpha$. For those, modal values of the hyperpriors are fixed resulting in $\alpha = 1/3$ for the DPM and $\alpha = 2/5$ for the dynamic MFM. Throughout this comparison, the sample size is fixed to $N = 100$.

\begin{figure}[t!]
  \centering
  \includegraphics[width=\textwidth]{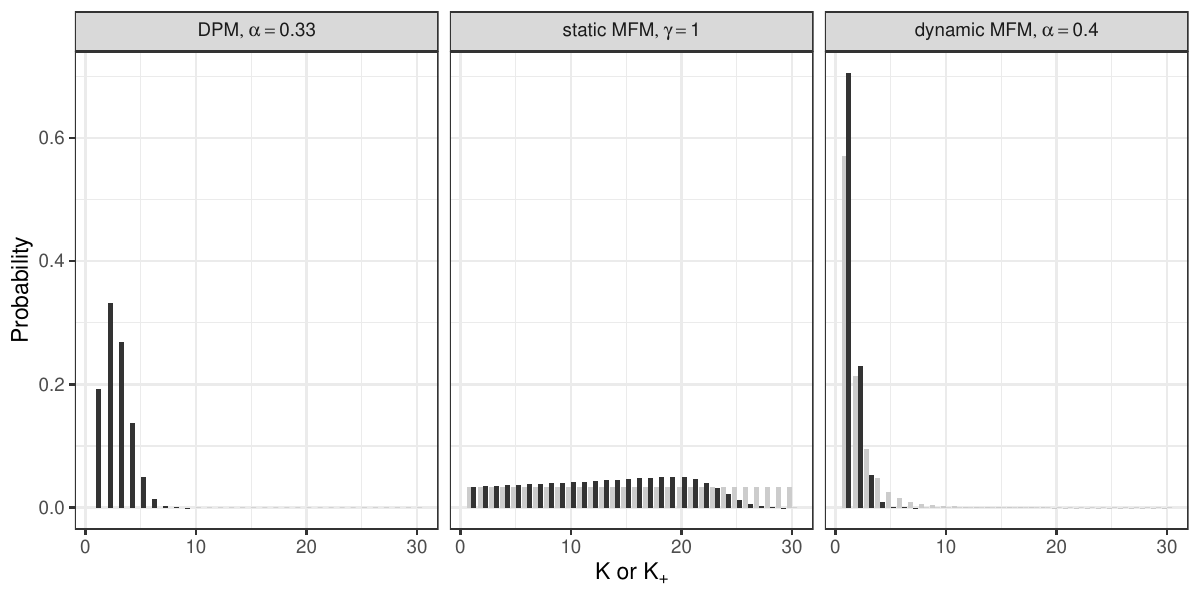}
  \caption{\label{fig:priorKplus}
    The prior probabilities of $K$ (in grey) and $K_+$ (in black) for
    the three modelling approaches.}
\end{figure}

Figure~\ref{fig:priorKplus}  shows the prior probabilities for $K$ and $K_+$ for the aforementioned three modelling approaches. For all three modelling approaches, clear differences between the imposed prior on $K$ and the implicitly obtained prior for $K_+$ are discernible.
The DPM approach with $\alpha = 1/3$ puts all mass at $K=\infty$ and
hence, only the implicit prior on $K_+$ is visualised. This prior is
unimodal with mode at $K_+ = 2$ and hardly any mass is assigned beyond
10. This implies that a sparse clustering solution with only a few
data clusters has high prior probability, but the homogeneity model is
not particularly supported a-priori. Similarly, as shown in
Section~\ref{sec:insp-induc-priors}, the dynamic MFM model with the
BNB$(1, 4, 3)$ prior on $K$ is also sparsity-inducing with high prior
mass concentrated around values ranging from 1 to 4 on $K_+$. However,
in contrast to the DPM approach, the homogeneity model is given by far
the highest prior probability. Finally, the static MFM with a uniform
prior $[1, 30]$ on $K$ behaves entirely differently as the differences
between the prior on $K$ and $K_+$ are smallest. Slightly increasing
probabilities for $K_+$ up to 20 indicate that a-priori no
penalisation to obtain a sparse solution is imposed in this setting.

\commentBG{Table~\ref{tab:results} characterises the prior on t}he
partitions implied by the three modelling approaches \commentBG{using
  some statistics to summarise} the induced prior on $K_+$ and the
balancedness of \commentF{the} partitions.
\begin{table}[h!]
  \begin{tabular}{l|c|c|c}
    \hline
    &DPM& static MFM & dynamic MFM \\
    \hline
    Mean of $K_+$& 2.6 &13.0&1.4\\
    Variance of $K_+$&1.5&45.5&0.4\\
    99\% quantile of $K_+$ &6&25&4\\
    Probability of $K_+ = 1$  (homogeneity)	&0.19&0.03&0.71\\	
    \hline		
    Relative entropy when $K_+=2$&0.45 (0.32)&0.73 (0.26)&0.51 (0.32)\\
    Relative entropy when $K_+=4$&0.57 (0.20)&0.79 (0.13)&0.59 (0.19)\\
    Relative entropy when $K_+=6$&0.64 (0.15)&0.82 (0.09)&0.65 (0.15)\\
    Relative entropy when $K_+=8$&0.68 (0.12)&0.84 (0.07)&0.69 (0.12)\\	
    Relative entropy (unconditional on $K_+$)&0.41 (0.25)&0.83 (0.14)&0.15 (0.14)\\
    \hline
    Number of singletons when $K_+=10$ &2.48 (1.28) &0.91 (0.87)&2.42 (1.27)\\
    \hline
  \end{tabular}
  \caption{Results regarding the characterisation of the induced prior on $K_+$ and the prior on the partition\commentBG{s} for the three modelling approach\commentBG{es}.
    For the relative entropy and the number of singletons,
    \commentBG{standard deviations are given in parentheses after the mean values}.}\label{tab:results}
\end{table}
\commentBG{Complementing the insights gained from}
Figure~\ref{fig:priorKplus}, the descriptive statistics
\commentBG{given in Table~\ref{tab:results}} to summarise the prior on
the number of data clusters $K_+$ for these three model specifications
also suggest that the specifications employed in the DPM and the
dynamic MFM are sparsity inducing while the static MFM has
\commentBG{a} more diffuse prior on $K_+$. \commentBG{The static MFM
  in fact has} \commentF{a} much higher variance and $99\%$ quantile
\commentBG{which is equal to} 25 with the upper bound of $K_+$
\commentBG{being in fact equal} to 30 for this model. Also, while the
DPM and the dynamic MFM both a-priori assume sparsity in the number of
data clusters, the prior probability assigned to \commentF{the}
homogeneity case is much higher \commentF{for} the dynamic MFM with
\commentF{a} probability of approximately 0.71 while that of the DPM
is about 0.19.

\commentBG{The results for the relative entropy indicate that t}he
static MFM with a uniform prior on $K$ \commentBG{gives} higher
probability a-priori to clusters with evenly sized blocks
\commentBG{than the other two specifications. This is} indicated by
\commentBG{the conditional and unconditional mean values of the}
relative entropy being much closer to 1 compared to \commentBG{the}
other \commentBG{two} models \commentBG{while the conditional and
  unconditional standard deviations are smaller}.
%
%
\commentBG{For} the DPM and the dynamic MFM, the relative entropy
conditional on $K_+$ seems to have almost identical mean and
corresponding standard deviation \commentBG{values} across all the
values of $K_+$ shown in the Table~\ref{tab:results}. However, the
unconditional relative entropy differs greatly between these two
specifications. \commentBG{Most likely this is due to} the difference
in the probability of homogeneity.

\commentBG{The results for} the number of singletons
\commentBG{indicate again that the prior mean and standard deviation
  are comparable for the DPM and the dynamic MFM specification
  conditional on $K_+ = 10$ with on average more than two clusters
  containing only a single observation. By contrast, only at most one
  cluster is a-priori expected to contain a single observation for the
  static MFM.}

Summarising the results of the comparison, it can be noted that the
DPM and the dynamic MFM \commentBG{using} the particular setting
employed in the previous literature induce a similar \commentBG{prior
  on the partitions conditional on the number of data clusters}
regarding the balancedness of the cluster sizes and the number of
singletons. Still, the latter approach is more sparsity inducing as
can be seen from the \commentBG{higher} prior probability of
homogeneity. In contrast, the static MFM employed with the uniform
prior on $K$ \commentBG{and on the weights assigns} much higher
\commentBG{weight} a-priori to clustering solutions
\commentBG{containing many data clusters and results in rather high
  prior mean values for the relative entropy}. This rather
counterintuitive result of fixing a diffuse prior on $K$
\commentBG{and the weights} resulting in an informative prior on
\commentBG{the} partitions should serve as a cautionary tale and
motivate the use of the proposed methodology to appropriately
characterise the induced prior on the partitions.

\section{Implications \commentF{for applied finite mixture analysis}}\label{sec:conclusions}

The modelling aims in model-based clustering depend on the specific
application and the prior domain knowledge available. Possible
scenarios are: (1) A coarse grouping of the data is aimed at in order
to identify basic structure, (2) a specific grouping is known and
should be reproduced without explicitly using this grouping in the
data analysis and (3) a flexible approximation of the data
distribution using many clusters is desired.

For the first case, clearly a sparsity inducing prior specification is
desired. This suggests to \commentF{use a} dynamic MFM with a prior on
$K$ with a mode at 1, decreasing probability weights for increasing
$K$ and a rather small value for $\alpha$ \commentBG{(see the specific
  dynamic MFM approach considered} in
Section~\ref{sec:comp-defa-priors}).

In the second case, the aim is to choose priors which induce implicit
priors such that the characteristics of the known grouping coincide
with the characteristics of the implicit priors. Clearly a suitable
prior \commentBG{specification} is characterised by inducing implicit
priors which assign substantial mass to known characteristics, such as
the number of data clusters $K_+$. If more detailed prior information
regarding the characteristics of the known partition is available,
\commentF{this} needs to be translated into some form of symmetric
additive functional \commentBG{where the prior mean and standard
  deviation may be computed given} the induced prior on the partitions.
Then 
$p(K)$ and $\gamma_K$ \commentF{can be chosen as
  to} 
match to their \commentF{desired} value. 
The suitability of such a specification would also need to be assessed
based on the sample size $N$.

In the third case, sparsity inducing priors are not desirable. The
prior specification, however, depends on the assumption of how complex
the approximation should be. E.g., the static MFM specification with a
uniform prior on $[1, 30]$ for $K$, $\gamma = 1$ and $N = 100$ assigns
rather comparable prior weights to values of the number of clusters
$K_+$ ranging from 1 to 20, thus encouraging approximations with many
clusters.  \citet{Richardson+Green:1997} use this specification in the
context of density approximation.
	

\section{Summary}\label{sec:summary}

\commentF{In this work, we} reviewed Bayesian cluster analysis methods
based on mixture models \commentF{and 
  explicit} priors imposed on the number of components and the weight
distributions for different modelling approaches. Fixing these
explicit prior specifications induces a specific, \commentF{implicit}
prior on the partitions. A thorough understanding of this particular
prior is of crucial interest in Bayesian cluster analysis. This is
because its \commentBG{characteristics} will in general be of
relevance when pursuing a specific modelling aim or assessing the
impact of specific prior combinations on the clustering result.  For
this reason, we derived computationally feasible formulas to
explicitly characterise the prior on the partitions. Specifically, to
serve this objective, the prior distribution on data clusters and the
first two moments of symmetric additive functionals computed over the
partitions both conditional and unconditional on the number of data
clusters are derived. Furthermore, the derivation of the formulas is
accompanied by a reference implementation in the \textsf{R} package
\textbf{fipp}.

\section*{Appendix}\label{sec:EPPF-support-combinatoric-part}
\subsection*{Combinatorial complexity of the prior enumeration}

By only considering cluster sizes as its argument, the EPPF introduced
in Section~\ref{sec:induced-eppf} simplifies the probability
assignment to all possible partitions of $[N] $ which otherwise will
equal the Bell number $B_N$. Furthermore, its symmetry enables all
sequences of cluster sizes that are equivalent under permutations to
be mapped into a unique sequence of \textit{ordered} data cluster
sizes
$(N_{(1)},N_{(2)}\ldots),\ N_{(1)}\geq N_{(2)}\geq \cdots > 0,\
\sum_{j} N_{(j)} = N$. Such a sequence of ordered data cluster sizes
is an element of the integer partitions of $N$. These two
simplifications combined give the total number of data cluster sizes
to be equal to the partition function of a positive integer $N$
written as $P_N$. For example, in $N = 4$ case, the value of $B_4$ is
15 while $P_4$ is 5. That is, for a sample size $N = 4$, even though
there are 15 unique partitions of $[N]$, only 5 distinct prior
probabilities are returned by the EPPF. In other words, the discrete
distribution of the induced prior partitions is summarised by the
equivalence classes introduced by the EPPF so that it only has 5
support points when $N = 4$.

Although this reduction of combinatorial complexity is substantial,
the support of the induced prior on partitions is still in the order
of $P_N$ which increases combinatorially with respect to an increase
in $N$. Even for a small sample size of $N = 100$, the $P_N$ is
approximately 190 million. Therefore, 
full enumeration of the EPPF is
computationally infeasible. Nor would a simple Monte Carlo approach be
able to adequately approximate the distribution, especially its higher
order moments.

  \label{sec:algor-comp-prior}
 \subsection*{Algorithm for computing \commentF{$C^{K, \gamma_K}_{N,k}$}}  \label{sec:algor-comp-prior}

  Algorithm~\ref{KNMFM} shows how to recursively determine
  $C^{K, \gamma_K}_{N,k}$ \commentF{defined in (\ref{PposCk})}.  $C^{K, \gamma_K}_{N,k}$ is required to
  determine the implicit prior on the number of data clusters and the
  conditional prior on the labelled data cluster sizes.
  \commentF{The recursion depends on a sequence of non-negative \lq\lq weights\rq\rq\ $\{w_n\}$}.
  For a static MFM, the weights $w_n$ do not vary for different number
  of components \commentF{$K$} and $C_{N,k}^{K, \gamma_K} \equiv C ^\gamma _{N,k}$ is
  independent of $K$. For a DPM, \commentF{with $K$ implicitly equal to $\infty$, $w_n = 1/n$ is even independent of $\alpha$}.
  To determine the prior $P(K_+ = k| N, \bm{\gamma})$ of the number of
  data clusters $K_+$, Algorithm~\ref{KNMFM} needs to be run once for
  static MFMs and for DPMs and consists of $N$ steps, i.e.,
  $\commentF{n}=1,\ldots,N$, while it needs to be run repeatedly for different
  values of $K$ for dynamic MFMs.

\begin{algorithm}[h!]
  \caption{Computing \commentF{$C^{K, \gamma_K}_{N,k}$} 
    for a general\commentBG{ised} MFM.}  \label{KNMFM}
  \begin{enumerate}
  \item Define the vector $\bm{c}_{K,1} \in \mathbb{R}^{N}$ and the
    $(N \times N)$ upper triangular Toeplitz matrix $ \bm{W}_1 $,
    where $ w_n = \frac{ \Gamma(n +\gamma_K )}{\Gamma(n +1)} $,
    $n=1, \ldots,N$,
\begin{align*}
\bm{W}_1  &= \left(
\begin{array}{cccc}
w_1 &    \ddots   &w_{N-1} & w_{N} \\
    &  w_1        &\ddots                 & w_{N - 1} \\
    &                 & \ddots                & \ddots  \\
    &              &                              &   w_1 \\
\end{array}
\right), & \bm{c}_{K,1}    &=   \left(   \begin{array}{l}  w_N  \\ w_{N-1} \\
\vdots    \\    w_1 \\ \end{array}   \right) .
\end{align*}
\item For all $k \ge 2$,  define  the  vector    $\bm{c}_{K,k} \in
\mathbb{R}^{N-k+1}$  as
\begin{align}  \label{recck}
\bm{c}_{K,k} &= \left( \begin{array}{cc} \bm{0}_{N-k+1 } &  \bm{W}_k    \\
\end{array} \right) \bm{c}_{K,k-1},
\end{align}
where $ \bm{W}_k $ is a $(N-k+1)\times (N-k+1)$ upper triangular
Toeplitz matrix obtained from $ \bm{W}_{k-1} $ by deleting the first row
and the first column.
\item Then, for all $k \ge 1$,
$C_{N,k} ^{K, \gamma_K} $ is equal to the first element of the vector $\bm{c}
_{K,k}$.
  \end{enumerate}
\end{algorithm}

\bibliographystyle{anzsj}
\bibliography{spying}


\end{document}